\documentclass[]{article}
\usepackage{graphics}
\usepackage{geometry}
\usepackage[section]{placeins}
\renewcommand{\vec}[1]{\mbox{\boldmath$#1$}}

\def\rot{\mathop{\rm rot}\nolimits}

\def\gsim{\lower.4ex\hbox{$\;\buildrel >\over{\scriptstyle\sim}\;$}} 
\def\lsim{\lower.4ex\hbox{$\;\buildrel <\over{\scriptstyle\sim}\;$}} 
\def\~  {$\sim$}

\def\beg{\begin{equation}}
\def\ende{\end{equation}}
\def\bega{\begin{eqnarray}}
\def\enda{\end{eqnarray}}
\def\pd{\partial}
\begin{document}

\title{The \( \alpha  \) effect and current helicity for fast sheared rotators:
some applications to the solar dynamo.}

   \author{V.V.Pipin$^1$}
\date{~}  
\maketitle
   \centerline{$^1$ \it Institute of Solar-Terrestrial
   Physics, Irkutsk}

   \abstract{We explore the $\alpha$ - effect
    and the small-scale current helicity,
   $h_c = \langle \vec h'\cdot {\rm rot}\ \vec h' \rangle$, for the case of
   weakly compressible magnetically driven turbulence that
   is subjected to the differential rotation. No restriction
   is applied to the amplitude of angular velocity, i.e.,
   the derivations presented are valid for an arbitrary
   Coriolis number, $\Omega^*=2\Omega\tau_{cor}$, though the
    differential rotation itself is assumed to be weak.
   
   The expressions obtained  are used to explore the possible distributions 
   of   $\alpha$ effect and $h_c$ in
   convection  zones (CZ) of the  solar-type stars. Generally, our theory 
   gives $\alpha^{\phi\phi}>0$ in the
   northern hemisphere of the Sun and the opposite case in
   the southern hemisphere. 
   In most cases the $h_c$ has the opposite sign to $\alpha^{\phi\phi}$. 
   However, we show that in
   the depth of CZ where the influence of  rotation upon
   turbulence (associated with $\Omega^*$) and the radial shear of angular
   velocity are strong,  the   distribution of
   $\alpha^{\phi\phi}$  might be drastically different from
   a classical $\cos\theta$ - dependence, where $\theta$ is
   colatitude.  It is shown that $\alpha^{\phi\phi}$  has
   a negative  sign at the bottom and below of CZ at
   mid latitudes. There, the distribution
   of $h_c$ is also different from  $\cos\theta$, but 
   it does not change its sign with the depth.
   
   Further, we briefly consider these quantities in the disk geometry. 
   The application of the
   developed theory to dynamos in the accretion disk  is more
   restrictive because they usually have a strong differential rotation,
      $\Bigl| \pd \log\Omega/\pd\log r\Bigr|  > 1$.
      
      }

\vspace{2 cm}
\section{Introduction}
It is generally
accepted that the $\alpha$ - effect  is  one of the
most important ingredients of the mean-field dynamo (Moffat
1978, Parker 1979, Krause \& R\"adler 1980).  Although this effect has
been known for a long time,  there is still some debate about
its existence if the
nonlinear back reaction of magnetic field is taken into
account (Moffat 1978, Vainshtein \& Cattaneo 1992,
R\"udiger \& Kitchatinov 1993, Cattaneo \& Hughes 1996,
Field et al. 1998, Brandenburg 2001, Ossendrijver et al.
2001).  
Here we shall not
discuss this issue, but  refer the reader to some recent
papers, especially those concerned with this question (Field et
al. 1998, Brandenburg 2001, Ossendrijver et al. 2001). 

It is well known that the basic dynamo mechanism that is responsible
for the generation the large-scale solar magnetic field is
the combined action of helical turbulent motions, $\alpha$
effect, and differential rotation. It is the so-called
$\alpha\Omega$ (or, more generally, $\alpha^2\Omega$) dynamo.
The question as to which extent the differential rotation
itself can be responsible for maintaining the $\alpha$ was  
discussed recently by Brandenburg (1999), R\"udiger \& Pipin
(2000) and R\"udiger et al.(2001). 

There are several reasons for incorporating  the differential
rotation in the
theory of the $\alpha$ effect:

1)In the
$\alpha\Omega$ - type dynamo, the most important component
of the $\alpha$ effect is its azimuthal component,
$\alpha^{\phi\phi}$. In standard theory, $\alpha^{\phi\phi}$ varies as
$\cos\theta$, with $\theta$ being colatitude (R\"udiger \&
Kitchatinov, 1993). This simple dependence is probably in
contradiction with observations because such an $\alpha$
can cause an intense magnetic field at the poles (cf. 
Brandenburg 1994, R\"udiger \& Brandenburg 1995). This
restriction is probably less important for the dynamo
operating in the whole CZ. Nevertheless, further indirect evidence for $\alpha$
having a maximum magnitude at low latitudes come from
examining the current helicity observations, see Pevtsov
et al (1995) and Kuzanyan
et al (2000). Their results indicate that current
helicity has perhaps a maximum at latitudes near $\sim
30^{\circ}$. Both the $h_c$ and the $\alpha$
effect might originate from a common source.
This source is well known. It is either stratified or
compressible turbulence that is
subjected to the Coriolis forces associated with
the shear flow (either rigid or differential
rotation). Rigid rotation gives 
$\alpha^{\phi\phi}\propto \cos\theta$. The influence of the 
differential rotation upon the turbulent convection
could give a more complicated latitudinal
dependence of $\alpha^{\phi\phi}$.

2) There are some arguments in
favor of the change of the sign of $\alpha$ due to the influence of
differential rotation if it is strong enough
(Brandenburg 1999, R\"udiger \& Pipin 2000).

Some of the first results on it were presented 
by R\"udiger \& Pipin (2000) and by R\"udiger et al.
(2001). The present paper generalizes those results for
the fast rotation case and it is important for
the astrophysical systems where the case $\Omega^* \ge 1$
is quite typical.

All  derivations in the paper are made for the case
of  weakly compressible magnetically driven
turbulence. The inclusion  of the small but
finite compressibility (in the sense that the density
fluctuations are allowed for) 
seems necessary  for an existence of
$\alpha$ in the {\it originally} homogeneous and isotropic turbulence
subjected to the influence of the mean sheared flow. It is
known that the density and the turbulence
intensity stratifications are probably the most important
contributions to $\alpha$ in the convection zones of 
late-type stars (R\"udiger \& Kitchatinov 1993).
 However, as a first approximations we
decided to investigate the influence of the differential
rotation on the $\alpha$ and $h_c$ in  "simpler" case and then to go ahead in case results 
 are promising.

Here we assume the turbulence to be
magnetically driven in the sense that the original (background)
turbulence consists
only of magnetic fluctuations and not of velocity fluctuations.
 This is probably a reasonable
assumption for  accretion discs where turbulence could be
induced by the magneto-rotational instability (Balbus \& Hawley 1991).  
The situation  in  convection zones 
of cool stars is different. In solar plasma the energy of 
the magnetic part of turbulent energy
is likely to be in equipartition with the hydrodynamic one 
(Krause \& R\"adler 1980, Vainshtein 1980, Biskamp 1997).
So, for reliable estimation of $\alpha$ it is necessary to
take 
both the "hydrodynamic" and the "magnetic" 
parts of turbulence simultaneously into account as was done,
for example, in the paper by Field et al. (1999). The
disadvantage of their computations is that
the spectrum of magnetic and hydrodynamic helicity in the 
background turbulence were prescribed a priori. 
We leave the derivation of $\alpha$ and $h_c$ with the "magnetic" and "hydrodynamic" parts
included simultaneously for the stratified differentially rotating flows for  a future 
on the reason said in absatz above. 
 
In this paper we derive both the $\alpha$ and $h_c$. 
One reason to do this is that the relation between these effects
has been commonly used as a diagnostic tool for $\alpha$-effect in solar physics
(Seehafer 1990, Pevtsov et al 1995, Kuzanyan et al 2000).
Another reason to consider both effects at a time is due to
the fact fact that  magnetic part of $\alpha$ is
often associated with the small-scale current helicity. 
 The opposite sign of
$\alpha$ and small-scale current helicity  that is
claimed by the Keinigs-Seehafer relations (Keinigs 1983;
R\"adler \& Seehafer 1990) can be understood from magnetic
helicity conservation law (e.g. Moffat 1978). The density of
magnetic  helicity has opposite signs on the large and small 
scales as a result of this conservation law. However, the sign of magnetic
helicity of the large-scale fields is the same as for $\alpha$
effect.

The paper is organized as follows, section 2 describes
the basic equations and approximations  we use in our
derivations. The general expressions for
$\alpha$ and $h_c$ are given. Section 3 is devoted to
examining the main effects in different situations. The applications to the
Sun are discussed in subsection 3.1. The
situation  in the disk geometry is
briefly considered in the subsection 3.2.
 Finally, the last section summarizes and
discusses  all the findings.  

\section{Basic equations}
\subsection{Mean-field electrodynamics}
As usual for the mean field MHD, we assume
the approximate scale 
separation (Moffatt 1978,2000; Krause\&R\"adler 1980).
Then the generation  of 
the large-scale magnetic field is
described by the following approximation of the
mean electromotive force (EMF) of fluctuating
fields,  
\begin{equation}
 {\cal E}_i = \alpha_{ij}\bar B_j+...,
\label{3}
\end{equation}
with 
\beg
\vec{\cal E} =  \langle {\bf u}' \times {\bf
h}'\rangle.
\label{1}
\ende
It is assumed that the large-scale magnetic
field is spatially homogeneous. The current helicity is
defined as
\[h_c=\langle \vec
h\cdot {\rm rot}\vec h' \rangle\]

All the derivations below are based on the first-order
smoothing approximation. This approximation is justified
if either the  magnetic Reynolds number, $R_m$, or the
Struhal number, ${\rm St}$, are much less than 1 
(Krause \& R\"adler 1980, Moffat 1978).
The first condition is not relevant for astrophysical
systems. Recently, Field et al (1999) reconsidered the
applicability of
${\rm St} << 1$ to the mean-field MHD. They pointed out the
experimentally observed fact that ${\rm St\sim 0.2-0.3}$ 
in the ordinary hydrodynamic turbulence (Pope 1994). In MHD
turbulence, the motions could be largely hydrodynamic in
character for a modest back reaction of magnetic field,
Pouquet et al (1976). Although such a ${\rm St}$ is not very
small,
we may take it to be a small parameter for pertubation
procedure. In addition, as
pointed out by Moffat (2000), ${\rm St}$ could be expected
rather small on the fast rotating astrophysical systems
because  the turbulence there "is more akin 
to a field of weakly interacting inertial waves 
whose frequencies  are of the order of the angular
velocity of the system".

The force field maintaining the turbulence  should be 
defined in
the {\em comoving} frame of  
reference. In this case it does not contain any 
information about  
the mean flow and its gradients, and we can safely
use the original turbulence concept, which is
widely accepted in most mean-field derivations
(Krause \& R\"adler 1980). This turbulence is supposed to exist  in
the absence of the mean-magnetic field and the mean flow. 
 Note, that our derivations is different at this
point from computations made by, e.g.,
Blackman(2000), who uses the concept of the
original turbulence, however, the equations describing
the evolution of the fluctuating fields are
written in an inertial frame of reference.
 
Thus we have to write at first the equations in 
the new coordinate system.
The mean flow is defined by
\beg
\bar U_i = W_{ij} \tilde x_j .
\label{4}
\ende
The tilded quantities   are defined in the  {\em rest} coordinate system. 
Then 
\beg
T_{ij}(t)=\exp\hat{W}t = \delta_{ij} + W_{ij} t +
{t^2\over 2} W_{il} W_{lj} + \dots .
\label{5}
\ende
The comoving coordinates are given by
\beg
x_i = T_{ij}^{-1} \tilde x_j.
\label{7}
\ende
The derivatives are transformed after
\beg
{\partial \over \partial \tilde x_i} = T_{ij}^{-1} {\partial
\over \partial x_j},
{\partial \over \partial \tilde t} = {\partial \over \partial t} -
W_{im} x_m {\partial \over \partial x_i}
\label{10}
\ende
so that the  velocity field behaves as
\beg
\tilde u_i = W_{im} T_{ml} x_l + T_{ij} u_j.
\label{13}
\ende
The induction equation in the rest frame of reference is
\beg
{\partial \tilde{\bf B} \over \partial \tilde t} = \rot\left\{\tilde{\bf u} \times
\tilde{\bf B} - \eta \rot \tilde{\bf B}\right\}
\label{14}
\ende
In the comoving coordinate system it  becomes
\beg
{\partial B_i \over \partial \tilde t} = (u_i B_j - u_j B_i)_{,j} +
\eta \Delta B_i - \eta t(W_{pl} + W_{lp}) B_{i,pl}
\label{15}
\ende
to the first order in $\hat W$.

A continuity equation has the form,
\beg
{\partial \rho'\over \partial t}=W_{im}x_m\frac{\partial \rho'}{\partial x_i}-
\bar{\rho}\ {\rm div}{\vec u'}.
\ende 
Starting from the equation of motion 
\beg
{\partial \tilde u_i \over \partial \tilde t} = - {1\over \rho}\ {\partial
 \over \partial \tilde x_i}\left( \tilde p + \frac{{\tilde B}^2}{2\mu_0}\right)+
\frac{\tilde B_j}{\mu_0\rho} {\partial \tilde B_i
 \over \partial \tilde x_j} + \nu \nabla^2\tilde u_i+\frac{\rho'}{\rho}{\vec g},\nonumber
\label{16}
\ende
one finds that
\begin{eqnarray}
&&\left({\partial \over \partial t} - \nu \nabla^2 \right)u'_i =
-  2 W_{il} u_l - 2t\nu  W_{ln}{\partial^2 u_i \over \partial x_n \partial x_l}
\nonumber \\
&&-{\left(\delta_{il}-t( W_{il}+ W_{li})\right)\over \rho} {\partial \over \partial x_l}
\left( p + \frac{ B^2}{2\mu_0}\right)\\
&&+ \frac{B_l}{\mu_0\rho}   {\partial B_i \over \partial x_l}+
\frac{\rho'}{\rho}{\vec g},\nonumber
\label{17}
\end{eqnarray}
where the acceleration, $\vec g$, includes
contributions due to gravity and the centrifugal
force. Dividing the magnetic field for the mean and fluctuating
parts,
\beg
B_i=\bar{B}_i+h^{(0)}_i+h'_i,\label{mag}
\ende
where the contribution to the fluctuating magnetic field
itself is made by the background magnetic
fluctuation, $h^{(0)}_i$, ("the original turbulence") and
$h'_i$ are magnetic fluctuations caused by the distortion
of the mean field, $\bar{B}_i$. The latter is governed by
the following  linearized equation:
\beg
{\partial h'_i \over \partial t} = (u'_i\bar{B}_j - u'_j\bar{B}_i)_{,j} +
\eta \Delta h'_i - \eta t(W_{pl} + W_{lp}) h'_{i,pl}.
\label{fluc_mag}
\ende
In the comoving frame of reference the fluctuating part of the velocity field
 satisfies the equation:
\begin{eqnarray}
&&\left({\partial \over \partial t} - \nu \nabla^2 \right)u'_i =
-  2 W_{il} u'_l - 2t\nu  W_{ln}{\partial^2 u'_i \over \partial x_n \partial x_l}
\\
&&-{\left(\delta_{il}-t( W_{il}+ W_{li})\right)\over \rho}
{\partial \over \partial x_l}
\left( p' + \frac{ (\vec{\bar{B}}\cdot\vec{h}^{(0)})}{\mu_0}\right)
+ \frac{\bar{B}_l}{\mu_0\rho}   {\partial h^{(0)}_i \over \partial x_l}+
\frac{\rho'}{\rho}{\vec g}.\nonumber
\label{fluc_u}
\end{eqnarray}
Next, we extract the solid body rotation part from the 
mean flow applying
$W_{ij}=\varepsilon_{ipj}\Omega_p+V_{ij}$,  where
the term $V_{ij}$  is responsible for
differential rotation. 

Upon  Fourier-transforming  and substituting the last
expression for  shear, $\hat W$, we write the induction
equation as 
\beg
(-i\omega + \eta k^2) \hat h'_i = i( {\vec k}\cdot \bar{\vec B})\hat u_i
- 2 i\eta V_{pl} k_p k_l {\partial \hat h'_i \over \partial \omega}
\label{hf}\nonumber
\ende 
and the momentum equation as 
\bega
&&(-i \omega +\nu k^2) \hat u'_i + 2(\varepsilon_{ipj}\Omega_p+ V_{ij} )\hat u'_j=
-\left(i k_i-V_{(ip)}k_p\frac{\pd}{\pd\omega}\right) \label{uf}\\
&&\times\left[\frac{C^2_{ac}\hat{\rho'}}
{\rho}+\frac{ (\vec{\bar{B}}\cdot \hat{\vec h}^{(0)})}{\mu_0\rho} \right]
-2 i\nu V_{pl} k_p k_l {\partial \hat u'_i \over \partial \omega}
+ \frac{i( {\vec k}\cdot \bar{\vec B})}{\mu_0\rho}\hat{h}_i^{(0)}+
\frac{\hat \rho'}{\rho}g_i, \nonumber
\enda
where we take into account the relation between density and pressure fluctuations,
$p'=C^2_{ac}\rho'$ with $C^2_{ac}$ being the sound speed in the turbulent medium,
$V_{(ip)}=V_{ip}+V_{pi}$. The compressibility effects are described by
continuity equation,
\beg
-i\omega{\hat \rho'}=-i\rho(\vec k\cdot\vec{\hat u}')-V_{im}k_i\frac{\pd\hat\rho'}
{\pd k_m}\label{rf}
\ende
Equations (\ref{hf},\ref{uf},\ref{rf}) can be solved using a perturbation 
procedure for the small parameters 
$max(u'^2,\displaystyle{\frac {h^{(0)\
2}}{\mu_0\rho}})/C^2_{ac}$ and shear, $V_{ij}$.  The first
parameter controls the  compressibility effects. Finite compressibility
should be taken into account to obtain the non-zero 
contributions to $\alpha$ effect. The density fluctuations described by (\ref{rf})
allows for both the Archimedian force and magnetic buoyancy.  The combined action
of these forces and the generalized Coriolis forces produce
non-zero contributions to
the EMF. Actually, the final result contains the factor
$H_c^{-1}=g/C^2_{ac}$.  It can be considered
a measure of the typical scale-height in the turbulent
medium in which the
compressibility effects are important. An estimation of $H_c$
for the solar convection zone gives $H_c\sim H_p$, with $H_p$ being the pressure scale
height. This explains the fact that $\alpha$ effect obtained 
is the same order of magnitude as
 $\alpha$ from density stratification (cf. R\"udiger \&
 Kitchatinov 1993). The whole perturbation procedure for the
 small but finite compressibility was described by Kitchatinov \& Pipin (1993).

The spectrum of background magnetic fluctuations 
is assumed to be
stationary and spatially homogeneous,
\bega
&&\langle h^{(0)}_i({\vec k},\omega) h^{(0)}_j ({\vec k}',
\omega')\rangle = 
{\hat
{\cal B}(k,\omega) \over 16\pi k^2} \pi_{ij}
\delta({\bf k +k'})\delta(\omega + \omega'),
\label{21}
\enda
where $\pi_{ij}= \delta_{ij} - k_i k_j/k^2$. Unfortunately
the expressions for the $\alpha$ effect and $h_c$ in terms
of integrals of spectral
functions are very difficult to manage. So, we have to pass to the
{\it mixing-length approximation} (MLT) in final results. To
do this, the procedure
proposed by Kitchatinov (1990) is used. We put 
$(-i\omega + \eta k^2)^{-1}=(-i\omega + \nu k^2)^{-1}=\tau_{cor}$, with
$\tau_{cor}$ being the typical correlation time of turbulence. The magnetic
spectrum is approximated by 
\beg
{\cal B}(k,\omega)\sim 2 \langle h^{(0)\ 2}\rangle\delta(k-l_{cor}^{-1})
\delta(\omega),
\ende
\beg
\langle h^{(0)\ 2}\rangle=\int_0^{\infty}\int_0^{\infty}
 {\cal B}(k,\omega)dk d\omega
\ende
Results obtained for $\alpha$ and $h_c$ are given in subsections below.

\subsection{The $\alpha$ effect.}
Even within the MLT approximation a general structure of the $\alpha$ effect
resulting from the shear contributions is rather complicated, 
\bega
&&\alpha ^{ij}=\Bigl[ e^{i}e^{j}\varepsilon ^{lmn}\bigl( f_{29}g_{m}V_{nl}
+f_{30}e^{k}e_{l}V_{nk}g_{m}+
( \vec{e}\cdot \vec{g}) f_{31}e_{l}V_{mn} \bigr) \\
&&+2f_{28}( \Omega \cdot \vec{g}) e^{i}e^{j}+ 
2f_{1}( \Omega \cdot \vec{g}) \delta ^{ij}+
\varepsilon ^{ijl}V_{lm}\bigl( f_{14}g^{m}-f_{13}e^{m}
( \vec{e}\cdot \vec{g}) \bigr) \nonumber\\
&&+f_{5}\varepsilon ^{lmj}V_{l}^{i}g_{m}
+ \delta ^{ij}\varepsilon ^{lmn}\bigl( -f_{2}g_{m}V_{nl} 
+f_{4}e^{k}e_{l}V_{nk}g_{m}+( \vec{e}\cdot \vec{g}) 
f_{3}e_{l}V_{mn}\bigr)\nonumber\\
&&+f_{10}\delta ^{il}\varepsilon ^{mnj}( \vec{e}\cdot \vec{g}) e_{m}V_{nl}
+f_{9}\varepsilon ^{lmi}e^{n}g_{l}e_{m}V_{n}^{j}
+f_{23}e^{i}\delta ^{jl}\varepsilon ^{mnk}e_{k}g_{n}V_{ml} \nonumber\\
&&+ f_{6}e^{l}\varepsilon ^{mnj}V_{l}^{i}e_{m}g_{n}
+f_{16}e^{j}\delta ^{il}\varepsilon ^{mnk}V_{kl}e_{m}g_{n}
+\varepsilon ^{lmn}g_{m}e_{n}
\bigl( f_{22}e^{i}V_{l}^{j}+f_{15}e^{j}V_{l}^{i}\bigr)\nonumber\\
&&+e^{j}\varepsilon ^{lmi}V_{mn}\bigl( f_{19}e^{n}g_{l}+f_{18}g^{n}e_{l}-
f_{17}( \vec{e}\cdot \vec{g}) e^{n}e_{l}\bigr)+
f_{20}e^{j}\varepsilon ^{lmi}( \vec{e}\cdot \vec{g}) V_{lm}\nonumber\\
&&+\delta ^{jn}\varepsilon ^{lmi}V_{mn}\bigl( f_{12}g_{l} 
+f_{11}( \vec{e}\cdot \vec{g}) e_{l}\bigr) 
+2f_{27}g^{i}\Omega ^{j} +2f_{25}g^{j}\Omega ^{i}\nonumber\\
&&-f_{26}(g^{i}e^{j}+g^{j}e^{i})\varepsilon ^{lmn}e_{l}V_{mn} 
+f_{24}e^{i}\varepsilon ^{lmj}g^{n}e_{l}V_{mn} 
+f_{21}g^{i}\varepsilon ^{lmj}e^{n}e_{l}V_{mn}\nonumber\\
&&+ \varepsilon ^{lmi}V_{m}^{j}\bigl( f_{8}g_{l}+
f_{7}(\vec{e}\cdot \vec{g})e_{l}\bigr) 
+  g^{j}\varepsilon ^{lmi}\bigl( f_{33}e^{n}e_{m}V_{ln}+
f_{32}V_{lm}\bigr) \Bigr] 
\frac{\tau ^{2}_{cor}\langle h^{(0)\ 2}\rangle }{\mu
_{0}\rho C^{2}_{ac}},\nonumber  
\enda
where all functions with indices are functions
of Coriolis number, $\Omega^*$. Expressions
for all of them can be found in the Appendix.
For the slow rotation case we have
\bega
&&\alpha ^{ij}=\Bigl[ \left( \frac{2g^{i}\Omega ^{j}}{15}-
\frac{( \vec{\Omega} \cdot \vec{g}) \delta ^{ij}}{5}-\frac{g^{j}\Omega ^{i}}{5}\right) 
\nonumber\\
&&+\frac{1}{5}\left( \frac{2}{3}\varepsilon ^{ilm}g^{j}V_{lm}-
\varepsilon ^{ilm}g_{l}V_{m}^{j}-\frac{2}{3}\varepsilon ^{ijl}g^{m}V_{lm}\right)\\
&&+\frac{\varepsilon ^{ilm}g_{l}}{15}\left( V_{m}^{j}+
\delta ^{jn}V_{mn}\right) -
\frac{1}{15}\varepsilon ^{ijl}g^{m}(V_{lm}+V_{ml})\Bigr] 
\frac{\tau ^{2}_{cor}\langle h^{(0)\ 2}\rangle }{\mu _{0}\rho C^{2}_{ac}}.\nonumber  
\enda
It should be noted that for this particular case, the solid body rotation
part of the \( \alpha  \)~-~effect, the upper string, can 
be  restored
from the terms contributed by shear, the remaining part of
the tensor.
The substitution \( V_{lm}=-\varepsilon _{lmp}\Omega
^{p} \) should be made to accomplish this step.

\subsection{The current helicity.}
The derivation of the current helicity was made with
an additional restriction, for the sake of simplicity, namely,
with the assumption that the azimuthal component of
the mean field is predominant. The
final expression for $h_c$ is,
\bega
&&h_c=\Bigl[ \bar{B}^{2} \bigl( 2\psi _{0}(\Omega \cdot \mathbf{g})+ 
\varepsilon ^{ijk}\bigl( \psi _{1}g_{j}V_{ki}
+\psi _{2}e^{l}e_{i}V_{jl}g_{k}+\psi _{3}(\mathbf{e}\cdot \mathbf{g})e_{i}V_{jk}\bigr)
 \bigr) \nonumber\\ 
&&+\varepsilon ^{ijk}\bar{B}^{l}\bar{B}_{k}\bigl( \psi _{6}g_{j}V_{li} 
+\psi _{7}(\mathbf{e}\cdot \mathbf{g})e_{i}V_{jl}+\psi _{9}g_{j}V_{il}\bigr)
-\psi _{5}\varepsilon ^{ijk}g_{j}\bar{B}^{l}\bar{B}_{i}e_{k}e^{m}V_{lm}\nonumber\\
&&+\psi _{4}\varepsilon ^{ijk}g_{j}\bar{B}^{l}\bar{B}_{i}V_{kl}
+\psi _{8}\varepsilon ^{ijk}\varepsilon ^{lmn}\bar{B}_{n}g_{i}\bar{B}_{j}e_{l}V_{mk}
\\
&&+ \varepsilon ^{ijk}\varepsilon ^{lmn}\bar{B}_{n}
\psi _{10}g_{j}\bar{B}_{k}e_{m}V_{li}\Bigr] 
\frac{\tau ^{3}_{cor}\langle h^{(0)\ 2}\rangle }{l^{2}_{cor}\mu_0\rho C^{2}_{ac}}. 
\nonumber
\enda
For the slow rotation case we obtain, 
\bega
&&h_c=\Bigl[ 
\frac{2}{15}\varepsilon ^{ijk}\bigl( \bar{B}_{i}g_{j}\bar{B}^{l}V_{kl}-
2g_{j}\bar{B}^{l}\bar{B}_{i}V_{kl}\\
&&-\bar{B}^{2}g_{j}V_{ki}\bigl) + 
\frac{2(\Omega \cdot \mathbf{g})}{5}\bar{B}^{2}\Bigr] 
\frac{\tau ^{3}_{cor}\langle h^{(0)\ 2}\rangle }
{l^{2}_{cor}\mu_0\rho C^{2}_{ac}}.\nonumber
\enda
Again the last term can be restored from the rest by the
method mentioned above. 
Both (24) and (26) are in agreement with previous
findings reported by R\"udiger \& Pipin (2000).

\section{Some applications.}
To find the expressions for $\alpha$ and $h_c$ for  a particular
coordinate system the back substitution $V_{ij}\rightarrow W_{ij}-\varepsilon_{ipj}\Omega_p$
has to be made, with $W_{ij}$ as a notation for the
derivative (the covariant one in the general case) of the large-scale
flow. Below, we examine  only the most important component of the
$\alpha$ effect, namely, $\alpha^{\phi\phi}$.

\subsection{The spherical geometry.}
For the case of the spherical geometry  we obtain,
\bega
&&\alpha ^{\phi \phi }=\Bigl[-\cos \theta f_{1}\Omega^*+
\frac{\cos \theta \sin ^{2}\theta }{2}\frac{\partial \log \Omega }
{\partial \log r}f_{\alpha 1}\\
&&+\frac{\sin \theta }{2}\frac{\partial \log \Omega }
{\partial \theta }\left( f_{\alpha 2}-
f_{\alpha 1}\sin ^{2}\theta \right)\Bigr]
\frac{g\tau_{cor}\langle h^{(0)\ 2}\rangle }
{\mu _{0}\rho C^{2}_{ac}}\nonumber \\
&&h_c=\Bigl[-\cos \theta\psi _{0}\Omega^* +
\frac{\cos \theta \sin ^{2}\theta }{2}\frac{\partial \log \Omega }{\partial \log r}\psi _{h1} 
\\
&&+\frac{\sin \theta }{2}\frac{\partial \log \Omega }{\partial \theta }
\left( \psi _{h2}-\sin ^{2}\theta \psi _{h1}\right)
\Bigr] 
\frac{g\tau ^{2}_{cor}\langle h^{(0)\ 2}\rangle
\bar{B}_{\phi}^2}
{l^{2}_{cor}\mu_0\rho C^{2}_{ac}},\nonumber
\enda
where $f_{\alpha 1}=(f_3-f_4-f_6-f_7+f_9)\Omega^*$, $f_{\alpha
2}=(f_3+f_5-f_2-f_7-f_8)\Omega^*$, $\psi
_{h1}=(\psi_2+\psi_3-\psi_5)\Omega^*$, $\psi
_{h2}=(\psi_1+\psi_3+\psi_6)\Omega^*$. Its expressions are given in
the Appendix.  In the slow-rotation limit we have
\bega
&&\alpha ^{\phi \phi }=\frac{2\Omega^*}{15}\left(
\frac{3\cos \theta}{4}+
\sin \theta\frac{\partial \log \Omega }
{\partial \theta } \right)\frac{g\tau_{cor}\langle h^{(0)\ 2}\rangle }
{\mu _{0}\rho C^{2}_{ac}} \label{sa_sp} \\
&&h_c =-\frac{\Omega^*}{5}\left(\cos \theta+\sin \theta
\frac{\partial \log \Omega }{\partial \theta }\right)
\frac{g\tau ^{2}_{cor}\langle h^{(0)\ 2}\rangle \bar{B}_{\phi}^2}
{l^{2}_{cor}\mu_0\rho C^{2}_{ac}}.\label{sh_sp}
\enda
For the SCZ ${\partial \log \Omega} / {\partial \theta} < 0.3$ and we can
conclude that in the slow rotation limit $\alpha ^{\phi \phi }>0$ and $h_c<0$ for 
the northern hemisphere, and they have opposite signs for the
southern hemisphere.

For the fast rotation case we obtain,
\bega
&&\alpha ^{\phi \phi }=\Bigl[\frac{\cos\theta}{2 \Omega^*}+
\frac{\pi\sin\theta}{128}\Bigl(\frac{9}{2}
\cos \theta \sin\theta
\frac{\partial \log \Omega }{\partial \log r}{\label{a_sp}}\\
&&+\frac{\partial \log \Omega }
{\partial \theta }\left( 7-\frac{9}{2}\sin ^{2}\theta \right)\Bigr)\Bigr]
\frac{g\tau_{cor}\langle h^{(0)\ 2}\rangle }
{\mu _{0}\rho C^{2}_{ac}},\nonumber \\
&&h_c= \Bigl[-\frac{\cos\theta}{\Omega^*}+
\frac{\pi\sin\theta}{64}\Bigl(3
\cos \theta \sin\theta
\frac{\partial \log \Omega }{\partial \log r}{\label{h_sp}}\\
&&-\frac{\partial \log \Omega }
{\partial \theta }\left( \frac{7}{2}+3\sin ^{2}\theta \right)\Bigr)\Bigr]
\frac{g\tau ^{2}_{cor}\langle h^{(0)\ 2}\rangle \bar{B}_{\phi}^2}
{l^{2}_{cor}\mu_{0}\rho C^{2}_{ac}}, \nonumber
\enda
where we retain the contributions like $\sim 1/\Omega^*$
in  zero order terms relative to the shear because
the latter is assumed small. For estimation
we put $\theta=\pi/4$. Near the bottom of the SCZ
where we might expect $\Omega^*\gg 1$, the
latitudinal shear of the angular velocity is much
less than the radial one. For the radial shear there 
we estimate $\cos\theta\sin^2\theta
\pd\log\Omega/\pd \log r \le 0.3$, see Figure 2 below. So
for $\alpha ^{\phi \phi }$, we can estimate  the
conditions on the Coriolis number for
equipartition between the solid body rotation part and terms
 contributed by shear,
\beg
 \frac{1}{\Omega^*_{\alpha}}=-\frac{9\pi}{256}
\frac{\pd\log\Omega}{\pd\log r}, \ \ \ \ \Omega^*_{\alpha}\sim 30. 
\ende
For $h_c$ we have,
\beg
\Omega^*_{\cal H}= \frac{3}{2}\Omega^*_{\alpha}
\ende
These conditions are probably too severe for the SCZ.
However the $\Omega^*_{\alpha}$ may
well be expected beneath
CZ, in the overshoot layer. Thus we can conclude that
the condition for sign reversal of the  $h_c$ (
{\it in the sense of the theory developed}) is hardly fulfilled on
the Sun. 
 
\vspace{0.3cm}
\begin{figure}[htbp]
{\centering
	\resizebox*{11cm}{7cm}{\includegraphics{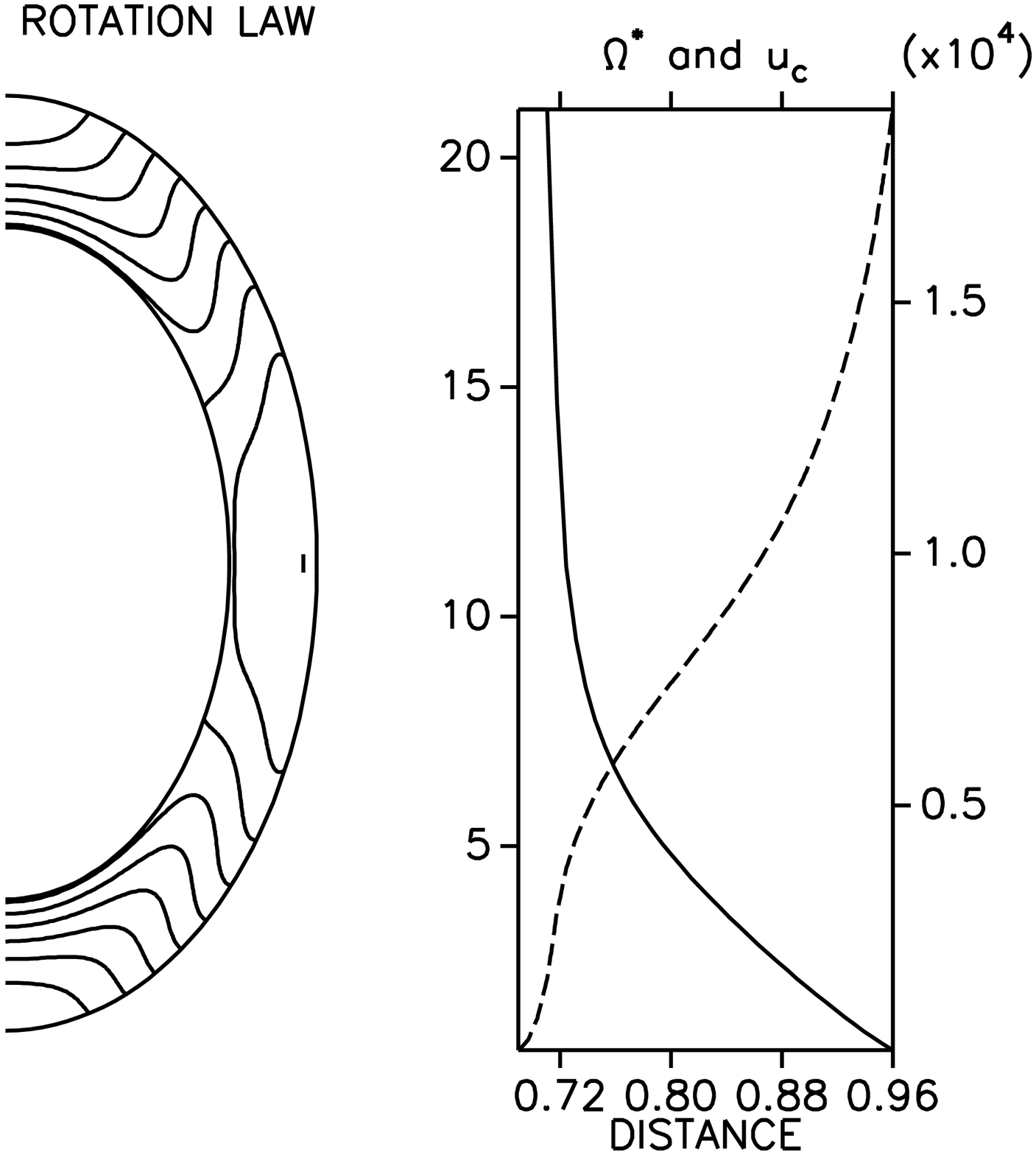}}\par}
\vspace{1cm}
{\centering
	\resizebox*{14cm}{7cm}{\includegraphics{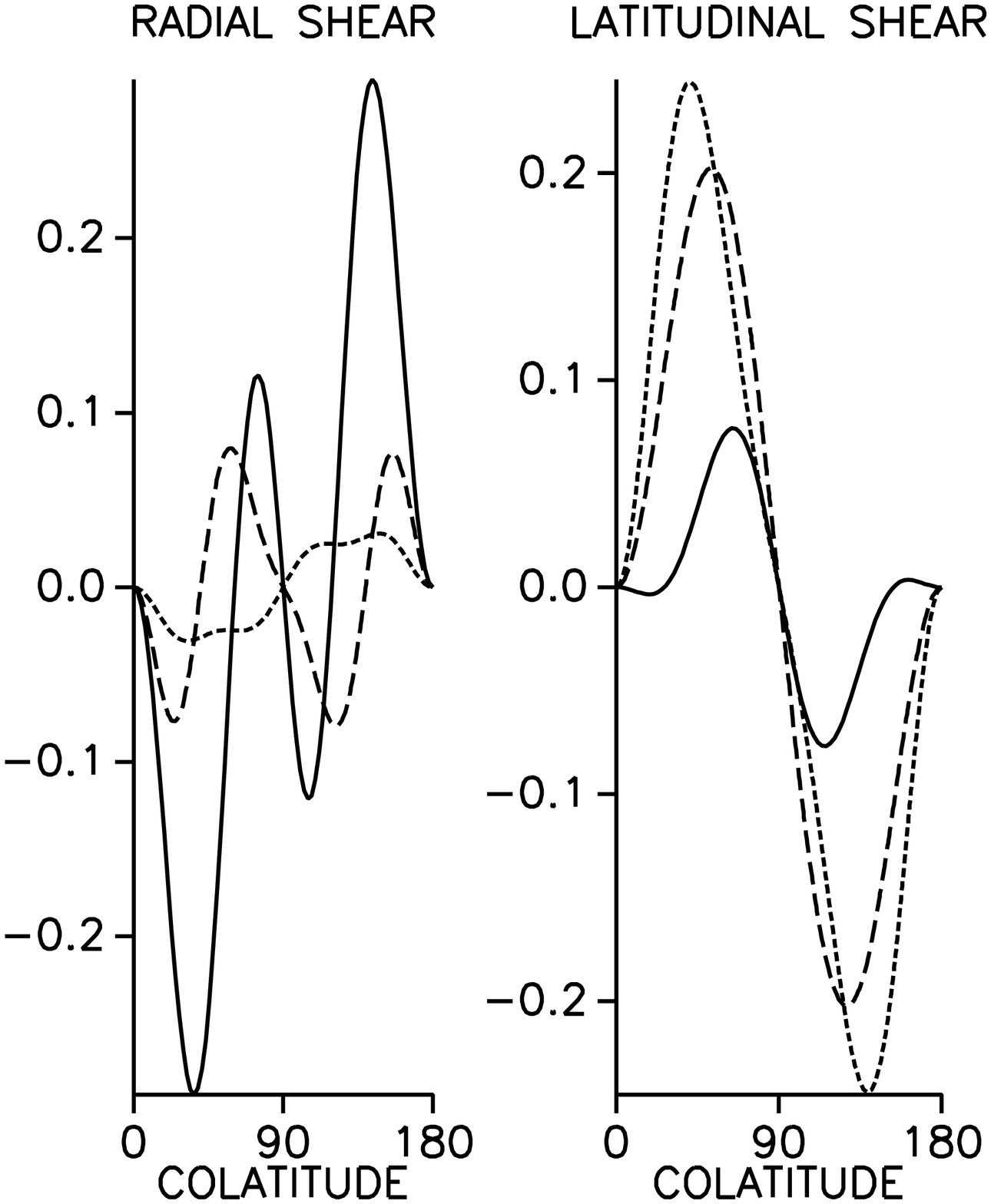}}\par}
	\caption{The rotation law is shown in the upper left panel. The
	 upper right panel
	shows the distribution of the Coriolis number, $\Omega^*$ (solid line),
	and the convective velocity, $u_c$ (dashed line),
	which is
	measured in $[cm/s]$. The lower panel
	shows the shear contributions. The term $\cos\theta\sin^2\theta
\pd\log\Omega/\pd \log r$ is shown at the lower left, and
the term $\sin\theta\pd\log\Omega/\pd\theta$ is shown at the
lower right.  The solid
lines are correspond to the bottom of the SCZ,
$0.715R_{\odot}$, the large dashed line to $0.84R_{\odot}$,
and the small dashed lines to  $0.95R_{\odot}$.}
\end{figure}
\vspace{0.3cm}
\begin{figure}[tp] 
 {\Large $\alpha^{\phi\phi}$} \\
 {\centering \resizebox*{10cm}{7cm}{\includegraphics{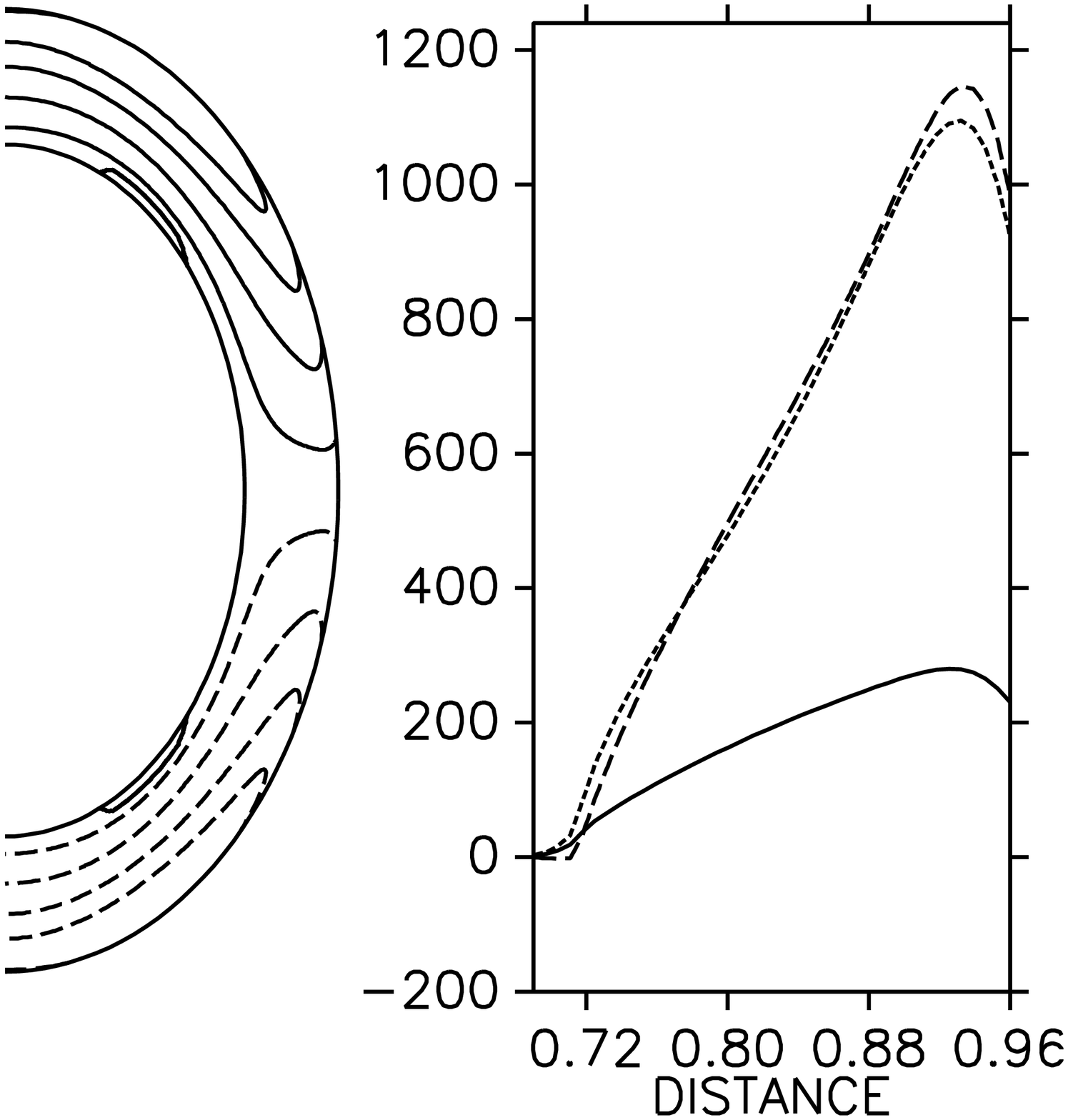}} \par}
{\Large $\alpha_{ff}$}  \\
{\centering 
 \resizebox*{10cm}{7cm}{\includegraphics{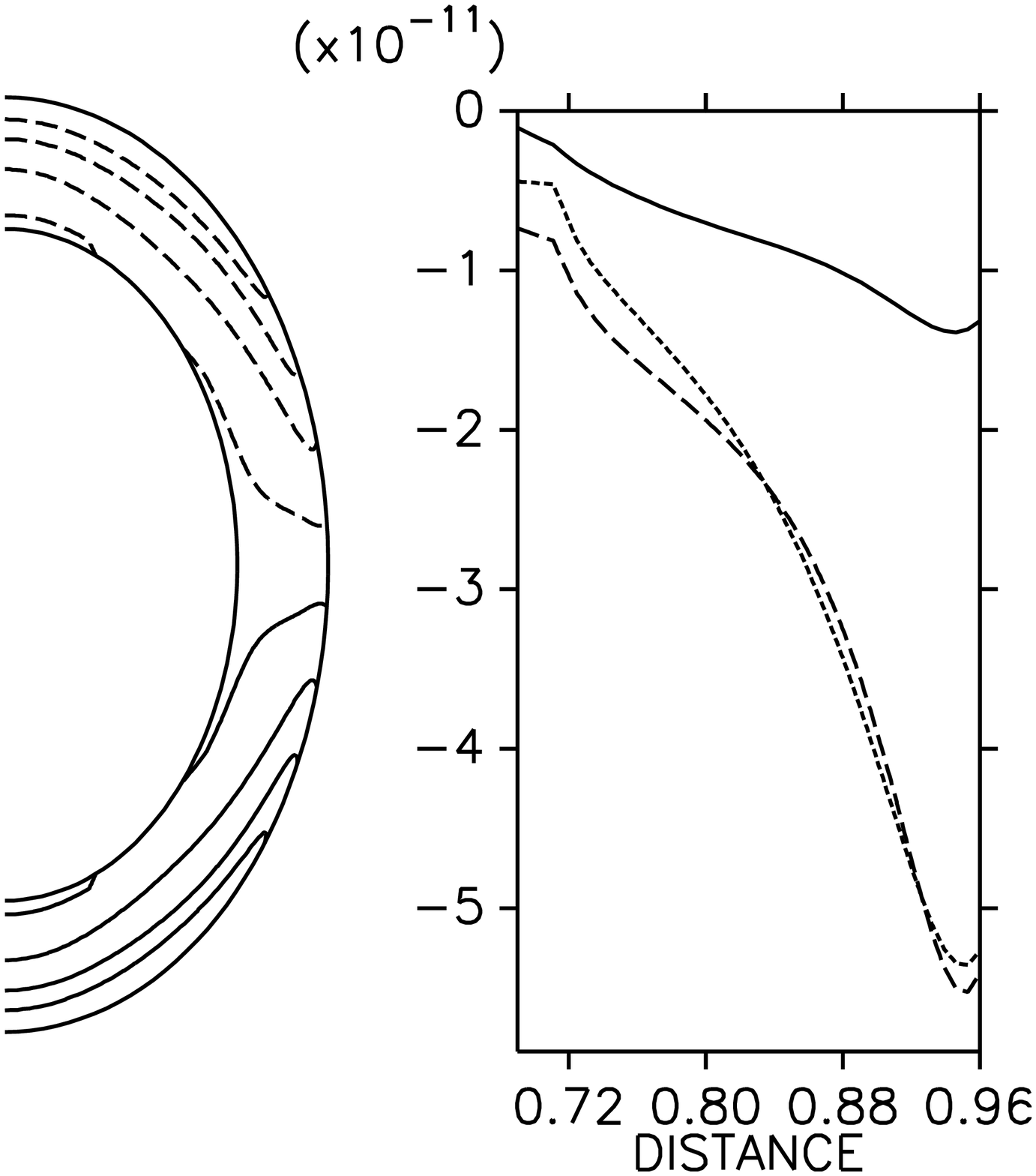}}\par}
\caption{ The resulting distribution of 
$\alpha^{\phi\phi}$ is shown in the upper panel and
$\alpha_{ff}$ is shown in
the lower panel. The left-hand parts shows the iso-contours in meridional
section. The Radial dependencies at different latitudes are shown at 
the right. The solid line corresponds to $15^{\circ}$ latitude
, the large dashed line corresponds to $45^{\circ}$ and the
small dashed line refers to the pole. The $\alpha^{\phi\phi}$ is 
measured in [cm/s], and $\alpha_{ff}$ is measured in [1/cm].}
\end{figure}	

To estimate the $\alpha ^{\phi \phi }$ and $h_c$ in SCZ 
more accurately we use the model of the solar
interior given by Stix (1990) and the helioseismology data
reported by Kosovichev et
al (1997)
with an analytical fit given by Belvedere et al.(2000).  The distribution of 
 $h_c$ depends on the spatial distribution of
 the large-scale magnetic field $\bar B_{\phi}$ which is
 known only hypothetically. Then we introduce the so-called
 "force-free alpha", $\alpha_{ff}=h_c /B^2_{\phi}$ (cf. Kuzanyan et al. 2000). Next, we use
 an assumption of the equipartition between the energy of the small-scale
  magnetic field and
the energy of the convective flows,
\beg
\frac{\langle h^{(0)\ 2}\rangle}
{\mu_{0}\rho}\approx  u_c^2,\ende
with  $u_c$ being the rms convective velocity. 
 We use a standard choice of the mixing-length parameter,
 $\alpha_{MLT}=1.6$. To explore the situation just beneath the bottom
 of the convection zone, $r_b=0.715R_{\odot}$, in the overshoot region,
  we use the analytical fit for the velocity there given by
 \beg
 u'_o=u'_b\left(\tanh(70(r-r_b))+1\right),
 \ende
 with $u'_b$ being the convective velocity at the bottom of
 the SCZ. The resulted depth of the overshoot layer is $\sim
 0.02R_{\odot}$.  The Coriolis number can be estimated by a procedure
 described by Kitchatinov et al.(2000). Inside the overshoot
 region, the Coriolis number was set equal to that at the bottom of the CZ. 
 The overshooting was followed to $0.69R_{\odot}$. Some important quantities
  for the SCZ , namely, the Coriolis number, the rms convective
 velocity and the shear contributions, $\cos\theta\sin^2\theta
\pd\log\Omega/\pd \log r$ and
$\sin\theta\pd\log\Omega/\pd\theta$ are presented in Figure 1.
The first inspection of Figure 1 shows that near the
bottom of SCZ the radial shear makes a positive contribution to the
$\alpha$ effect at the near equatorial latitude and a
negative contribution at  mid and high latitudes.  The same can
be said about $h_c$. This 
explains the results presented on the Figures 2 and 3. Figure
2 shows the general distributions of the
$\alpha^{\phi\phi}$ and  $h_c$ in
the meridional section(left panels) and the 
radial profiles of these quantities at different latitudes. 

\begin{figure}[tbp]
{\centering 
 \resizebox*{15cm}{7cm}{ \includegraphics{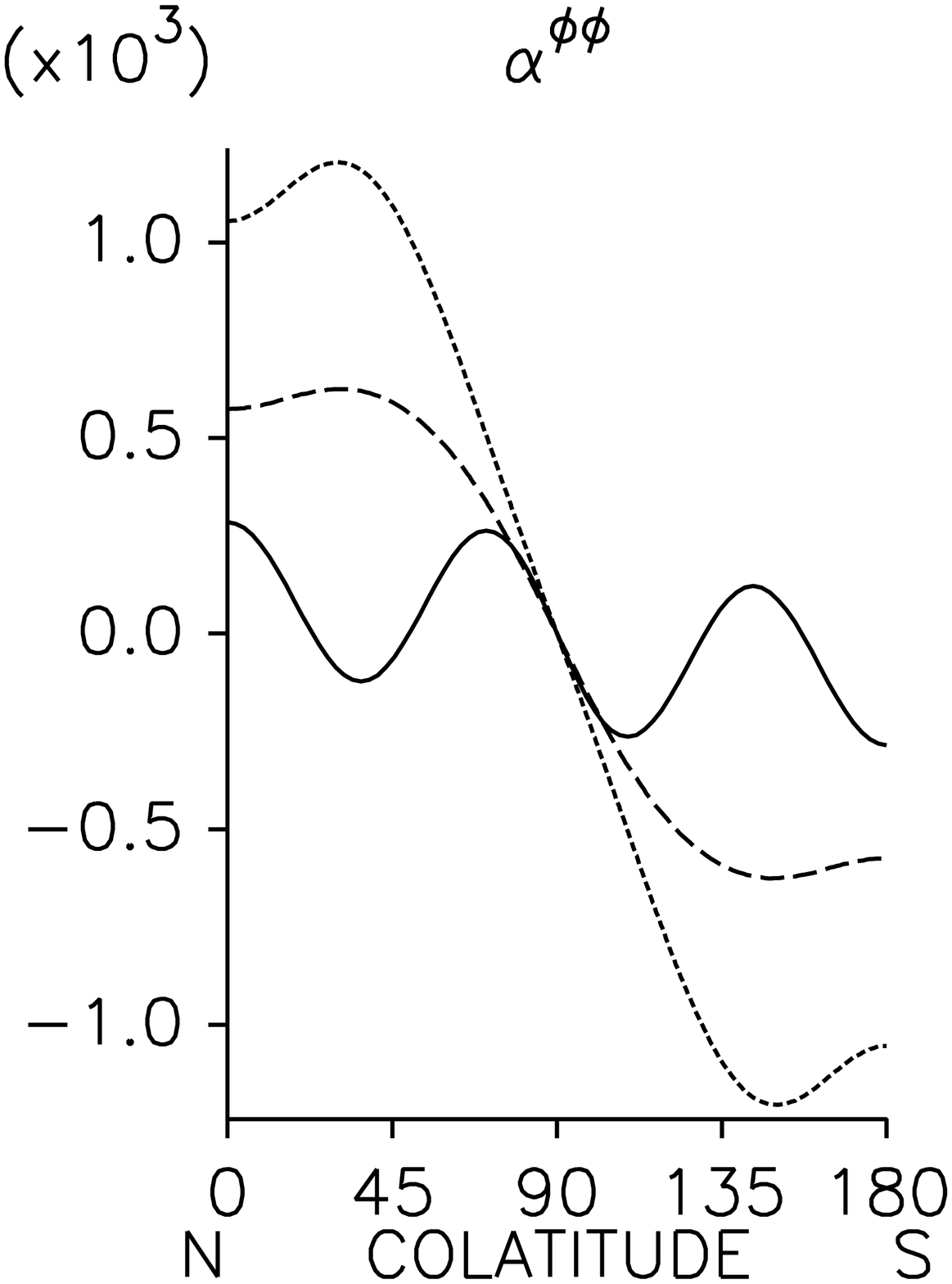}
 \includegraphics{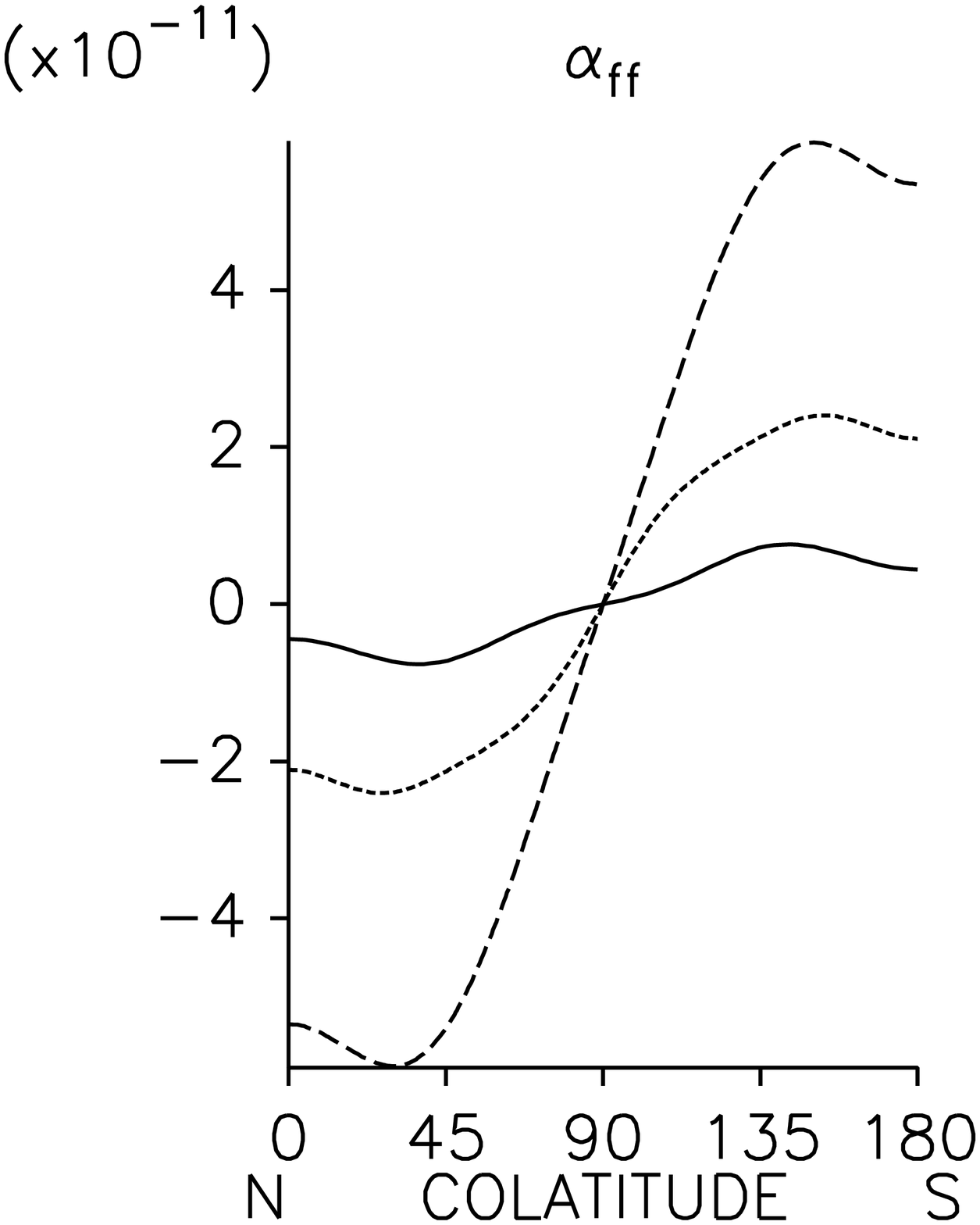} } \par}
\caption{Latitudinal distributions of 
$\alpha^{\phi\phi}$ (left), and 
$\alpha_{ff}$ (right), respectively. Solid lines show the
values
at the bottom of the SCZ, $0.715R_{\odot}$. We choose to
multiply $\alpha^{\phi\phi}$ by factor 100 at this level to
present all the dependencies at a time.
Large dashed lines show the dependence at the level of
$0.84R_{\odot}$,
and the small dashed lines correspond to the level of
$0.95R_{\odot}$. As in the previous Figure, the $\alpha^{\phi\phi}$ is 
in [cm/s] and $\alpha_{ff}$ is in [1/cm].}
\end{figure} 
The latitudinal distributions at different depths are shown
by Figure 3. A
complicated latitudinal dependence of $\alpha^{\phi\phi}$ at
the bottom of SCZ is evident. Note, that the obtained value of
$\alpha_{ff}$ is on the order
of magnitude in good agreement with observation. This
parameter was  computed from a different
set of the solar magnetograms by several authors (Pevtsov et
al. 1995, Kuzanyan et al. 2000).
 All of them estimated $\alpha_{ff}$ at $\sim 10^{-8}[1/m]$.
 
\subsection{$\alpha^{\phi\phi}$ and $h_c$ for disks.}
In disks, where the angular velocity  is dependent on the
radius, we obtain the following expressions for the
azimuthal component of $\alpha$ and $h_c$, 
\bega
&&\alpha ^{\phi \phi }=g_z\Bigl[f_{1}\Omega^*-
\frac{f_{\alpha2}}{2}\frac{\pd \log \Omega}{\pd\log r}\Bigr]
\frac{\tau_{cor}\langle h^{(0)\ 2}\rangle }
{\mu _{0}\rho C^{2}_{ac}},\\
&&h_c=g_z\Bigl[\psi_{0}\Omega^*-
\frac{\psi_{h2}}{2}\frac{\pd \log \Omega}{\pd\log r}\Bigr]
\frac{\tau ^{2}_{cor}\langle h^{(0)\ 2}\rangle \bar{B}_{\phi}^2}
{l^{2}_{cor}\mu_0\rho C^{2}_{ac}}. 
\enda
In the slow rotation limit we reproduce the results reported by
R\"udiger \& Pipin (2000), 
\bega
&&\alpha ^{\phi \phi }=-\frac{2g_z\Omega^*}{15}\Bigl[\frac{3}{4}+
\frac{\pd \log \Omega}{\pd\log r}\Bigr]
\frac{\tau_{cor}\langle h^{(0)\ 2}\rangle }
{\mu _{0}\rho C^{2}_{ac}},\label{sa_d}\\
&&h_c=\frac{g_z\Omega^*}{5}\Bigl[1+
\frac{\pd \log \Omega}{\pd\log r}\Bigr]
\frac{\tau ^{2}_{cor}\langle h^{(0)\ 2}\rangle \bar{B}_{\phi}^2}
{l^{2}_{cor}\mu_0\rho C^{2}_{ac}}.\label{sh_d}
\enda
In the fast rotation case we have
\bega
&&\alpha ^{\phi \phi }=-g_z\left(\frac{1}{2\Omega^*}+
\frac{7\pi}{128}\frac{\pd \log \Omega}{\pd\log r}\right)
\frac{\tau_{cor}\langle h^{(0)\ 2}\rangle }
{\mu _{0}\rho C^{2}_{ac}},{\label{a_d}}\\
&&h_c=g_z\left(\frac{1}{\Omega^*}+
\frac{7\pi}{64}\frac{\pd \log \Omega}{\pd\log r}\right)
\frac{\tau ^{2}_{cor}\langle h^{(0)\ 2}\rangle \bar{B}_{\phi}^2}
{l^{2}_{cor}\mu_0\rho C^{2}_{ac}}{\label{h_d}}
\enda
\vspace{0.3cm}
\begin{figure}[htbp]
{\centering
	\resizebox*{13cm}{7cm}{\includegraphics{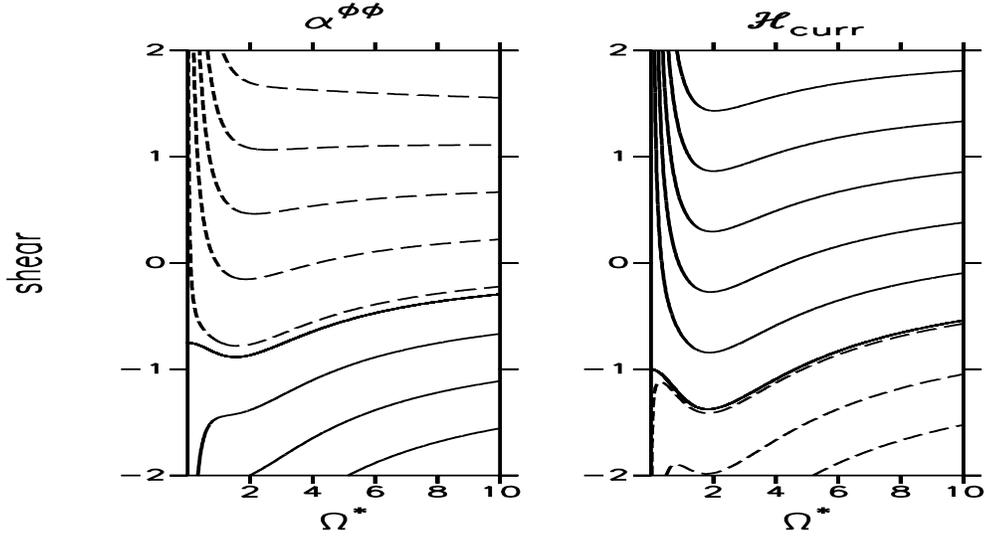}}\par}
\vspace{0.3cm}
\caption{The contours for functions controlling the signs
of $\alpha^{\phi\phi}$ and $h_c$, equations
(\ref{ad},\ref{hd}). Negative values are shown by the dashed lines.
 The effects under investigation have opposite signs almost
everywhere except the region $S\sim -1$.}
\end{figure}

In order to explore the relation between $\alpha^{\phi\phi}$ and 
${\cal {H}}_{curr}$ with due regard for the different Coriolis
numbers and the shear ${\pd \log \Omega/\pd\log r}$ we
construct two functions that define the sign of the effects
in the upper plane of the disk where $g_z < 0$. Namely, the function
\beg
\alpha_D=f_{1}\Omega^*-
\frac{f_{\alpha2}}{2} S, \label{ad}
\ende
where $S={\pd \log \Omega/\pd\log r}$ defines the sign of
the $\alpha^{\phi\phi}$, and so does the function
\beg
{\cal {H}}_D=\psi_{0}\Omega^*-
\frac{\psi_{h2}}{2}S,\label{hd}
\ende
for $h_c$. Figure 4 shows the
iso-contours for $\alpha_D$ and ${\cal {H}}_D$ for the different
$\Omega^*$ and $S$. From Figure 4 we conclude that the signs of
the $\alpha^{\phi\phi}$ and 
$h_c$ are opposite nearly for all values of
the Coriolis number and the shear except the region with
shear $S\sim -1$. Thus, the Keinig-Seehafer's identity
works well for the disks. However, it should
be pointed out that application of the theory developed here
is questionable for the fast rotating disks with $|S|>1$.
In our derivations we assumed that the differential rotation
is weak. Though, the applications for
the slow rotation results (40,41), and  for $|S|>1$
could be justified in our approximations.

\subsection{The dynamo with the shear-dominated $\alpha$
effect.}
This subsection is devoted to some preliminary
results of applications the obtained
$\alpha$ - effect to 1D $\alpha\Omega$ dynamo
model. The considered model is similar to those in
papers by Kitchatinov et al.(1994) and Kitchatinov \&
Pipin(1998). The detailed derivations
can be found there. Here, we only rewrite the
mathematical formulation of the problem with taking into
account new contribution to $\alpha$  effect due
to the shear.  
The system of the simplified 1D dynamo  equations
 with regards for the non-linearity caused by the
 influence of the large-scale magnetic field on
 the $\Lambda$ effect driving the differential
 rotation is following, 
\begin{eqnarray}
&&\frac{\partial \widetilde{\Omega }}{\partial t}=
\frac{Pm}{\sin ^{3}\theta }\frac{\partial }{\partial \theta }
\left[ \sin ^{2}\theta \left( \sin ^{3}\theta 
\frac{\partial \widetilde{\Omega }}{\partial \theta }-
\Omega _{0}\cos \theta \widetilde{H}\right) \right],
\nonumber \\
&&\frac{\partial A}{\partial t}=\psi _{\alpha
}\tilde{\alpha}B+\frac{\partial }{\partial
\theta }\left[ \frac{1}{\sin \theta
}\frac{\partial }{\partial \theta }\sin \theta
A\right], \label{sh} \label{1d} \\
&&\frac{\partial B}{\partial t}={\cal
{D}}\widetilde{\Omega }(\theta )\frac{\partial
\left[ \sin \theta A\right] }{\partial \theta
}+\frac{\partial }{\partial \theta }\left[
\frac{1}{\sin \theta }\frac{\partial }{\partial
\theta }\sin \theta B\right].\nonumber 
\end{eqnarray}
where $B$ is the toroidal magnetic field, $A$ is
the potential the poloidal component of the
field and $\tilde\Omega=\pd\log\Omega/\pd r$ is the radial
gradient of the angular velocity. The coefficient,
\beg
\tilde{\alpha}=\cos \theta \left\{1 +C_{sh}\,
\widetilde{\Omega }(\theta )\cos \theta \sin
^{2}\theta \right\},\label{ash}
\ende
where
\[
C_{sh}=C_1\frac{f_{\alpha 1}(\Omega^*)}{\Omega^*f_{1}(\Omega^*)}
\]
\begin{figure}[]
{\centering
	\resizebox*{8cm}{8cm}{\includegraphics{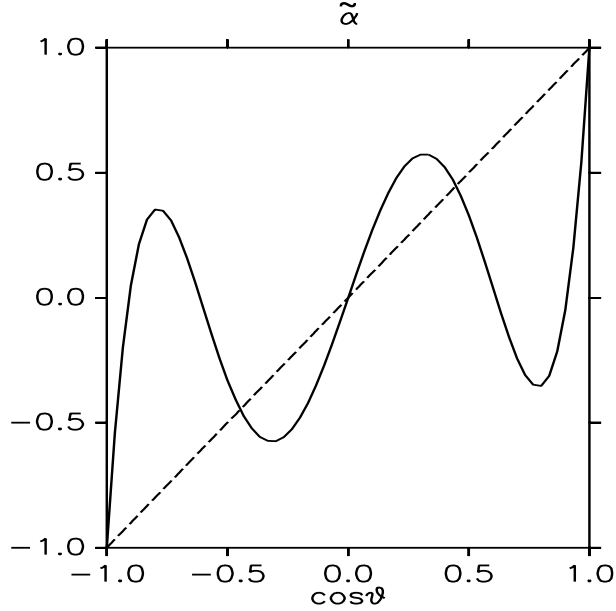}}\par}
\vspace{0.3cm}
\caption{The latitude dependence of $\tilde{\alpha}$ used
in the model. The dashed line shows the "standard"
dependence, $\alpha\propto\cos\theta$. The solid line shows
the shear-dominated $\alpha$, (\ref{ash}), with
$\Omega^*=20$ and $C_1=3$ (cf. Figure 3).}
\end{figure}
defines the distribution of the alpha effect. We introduced
another coefficient $C_1$ to match the distribution of  the alpha effect near the
bottom of the SCZ found in the
subsection 3.1.
The system (\ref{1d}) can
be considered as a simplified version of the
so-called
"interface dynamo" which might be operating at the interface
between CZ and the overshoot region (cf Parker 1993,
Charbonneau \& MacGregor 1997). 
The magnetic field is measured in
units of the field corresponding to equipartition
between the magnetic energy and the mean energy of
the convective flows, the shear, $\tilde\Omega$, is measured in
units of $\Omega_0/R_{\odot}$, and time - in units of
the typical diffusion time, $R^2_{\odot}/\eta_T$.
$P_m=\nu_T/\eta_T$ is a magnetic Prandtl number and
we shall assume $P_m=1$ in what follows, and
\[
{\cal D}=-\frac{\Omega\alpha R^3_{\odot}}{\eta_T^2}
\]
is the dynamo number. Functions 
\begin{eqnarray*}
&&\psi _{\alpha }=\frac{15}{32B^{4}}\left[
1-\frac{4B^{2}}{3\left( 1+B^{2}\right)
^{2}}-\frac{1-B^{2}}{B}\arctan B\right] ,  \\
&&\Phi _{0}=\frac{4}{B^{2}}\left[ \frac{2+3B^{2}}{2\sqrt{\left( 1+B^{2}\right) ^{3}}}-1\right] , \\
&&\Phi _{1}=\frac{2}{B^{2}}\left[ 1-\frac{1}{\sqrt{\left( 1+B^{2}\right) }}\right] . 
\end{eqnarray*}
describe the  magnetic feedback reaction upon the $\alpha$- and
$\Lambda$- effects.

If the influence of the magnetic field is neglected
then equations (\ref{1d}) yields a shear distribution
\beg
\widetilde{\Omega }=
\frac{1}{10}\left( 5\sin ^{2}\theta -4\right)\label{shd}.\ende
One has a negative radial angular velocity
gradient at the high latitudes and a positive
gradient at the equator. This is in agreement with
the helioseismology data. 

The distribution of the $\alpha$
-effect in (\ref{1d})  is described by $\tilde{\alpha}$. It
depends on the strength of the shear and the Coriolis
number. The model in subsection 3.1 gives
$\Omega^*=20$ near the bottom of the SCZ.
With the shear (\ref{shd}) and $C_1=3$ we can match the
distribution of the $\alpha^{\phi\phi}$ near the bottom of the SCZ
found in subsection 3.1. Next picture, Figure 5, shows the
dependence of $\tilde{\alpha}$ used in simulations. The dependence
$\alpha\propto\cos\theta$ is shown by
the dashed line. The solid line shows the  $\alpha$
contributed by shear. At the middle latitudes the difference
 between both alphas is evident. 

The system (\ref{sh}) was solved numerically. The initial
field was  weak ($B\ll 1$), and
did not has any particular symmetry with respect
to equator. We followed the field dynamics  to a steady-state regime
 that does not
depend on the initial field. Only such terminal solutions
will be considered later. 
\vspace{0.3cm}
\begin{figure}[htbp]
{\centering
\resizebox*{13cm}{19cm}
{\includegraphics{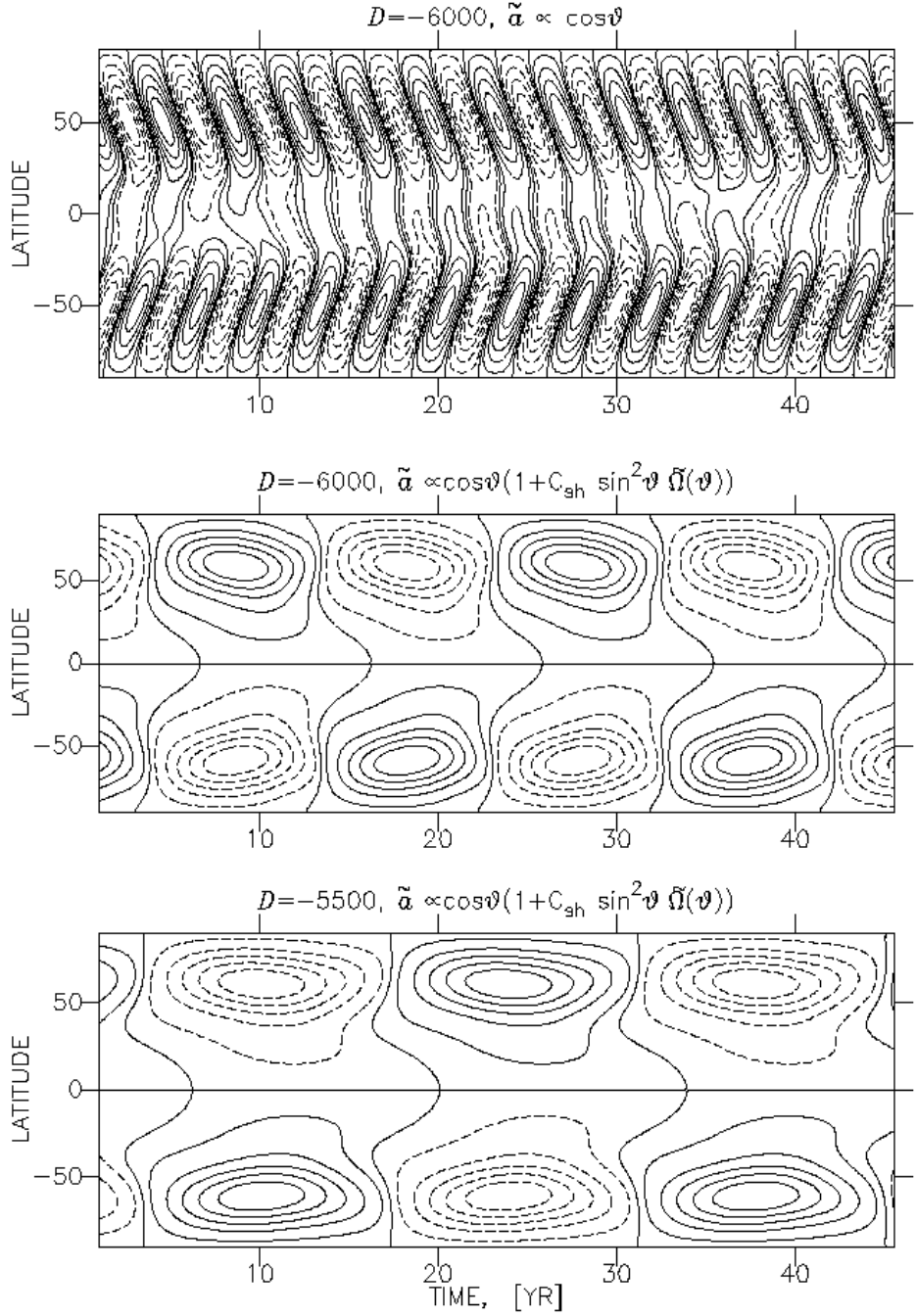}} \par}
\caption{The butterfly diagrams of the large-scale toroidal magnetic
field evolution for different type of the $\alpha$
effect distribution. Time was computed by taking
$\eta_T=10^{13}\ cm^2/s$ and $R_{\odot}=6.96\cdot10^{10}\ cm$
into account.}
\end{figure}
\vspace{0.3cm}

The critical dynamo-numbers and some results about the
non-linear behavior of (\ref{1d}) with $\alpha\propto \cos\theta$
 were considered in more detail by Kitchatinov et al.(1994) and Kitchatinov \&
Pipin(1998). Equations (\ref{1d}) give 
oscillatory solutions for negative dynamo numbers. The most
interesting solution with long-term modulations of magnetic
field and the shear was found by Kitchatinov et al.(1994) for a slightly
supercritical regime, ${\cal D}=-6000$. Here we repeat these
simulations for a different type of the $\alpha$ - effect.
  
Figure 6 shows some results of the numerical simulations.
The upper panel demonstrates the long-term evolution of the
butterfly diagrams of the toroidal field in coordinates 
time-latitude for $\alpha\propto \cos\theta$. This diagram
is in agreement with the previous findings in papers by
Kitchatinov  et al.(1994) and Kitchatinov \&
Pipin(1998). The middle panel on Figure 6 shows the same for an
$\alpha$ effect dependent on  the shear, (\ref{ash}), and with the same
${\cal D}=-6000$. There are several differences between
pictures. The first one
is that the second case does not show the long-term
oscillations despite the
non-linear interplay between magnetic field and shear.
Next, the butterfly wings are too concentrated near the pole. The
last difference is that period of the considered dynamo
is about 5 times larger. This effect is likely due to the
space separation between $\alpha$ and $\Omega$ effects on
latitude. Such an effect was discussed firstly by Deinzer et al.
(1974).  This latitude separation is also probable reason
for the polar concentration of magnetic field. The down
panel on Figure 6 shows the evolution of the toroidal
magnetic field with the general type of $\tilde\alpha$,
(\ref{ash}), and with  ${\cal D}=-5500$. The period of
dynamo is 8 times large there than for  the "usual" $\alpha$
effect and it is comparable with the solar cycle period.

The critical dynamo number for the dynamo with the shear-dominated
$\alpha$ effect is about ${\cal D}=-5400$. We also tried to
explore  
the limiting case where the $\alpha$ was defined only by the
shear contributions, i.e, $\tilde\alpha$ has form like 
$\tilde\alpha\propto\cos\theta\sin^2\theta\tilde\Omega$.
Only steady solutions were found for the negative dynamo
numbers in this case. This issue is also in agreement with analysis
in the paper by Deinzer et al. (1974), who also found the
steady type solution after the space separation between 
$\alpha$  and $\Omega$ effects approaches to some limit.

\section{Conclusions.}
   Let us summarize our findings. We have explored the $\alpha$ effect
    and the small-scale current helicity for the case of 
   weakly compressible magnetically driven turbulence that
   is subjected to differential rotation. The derivations
   presented here were made without restriction on
   the  angular velocity amplitude, i.e., for an arbitrary
   Coriolis number. The differential rotation itself
   assumed to be weak. The main finding is that for the fast
   rotation case the influence of the differential rotation
   on the $\alpha$ effect and $h_c$ can be quite
   strong. Both $\alpha$ and $h_c$ become
   non-monotonic functions of the latitude with two maxima in
   one hemisphere. One maximum is at mid
   latitude and another is at the pole as in
   standard theory. 
   
   Beneath the solar convection zone, at the northern
   hemisphere,  $\alpha^{\phi\phi}$ is likely to have a negative sign
   at mid-latitudes. However, this result is not
   conclusive. The problem is that for the magnetically driven
   $\alpha$ effect the "standard", $\cos\theta$-like
   contributions quenches for the fast rotation case as
   $\Omega^{*\ -1}$, and the shear contributions approaches to
   a constant. Thus the negative sign of the
   $\alpha$ is the result of such a behavior of the competitive
   contributions. Since we know the hydrodynamic part of the
   $\alpha^{\phi\phi}$ approaching a constant
   under the fast rotating limit (R\"udiger \& Kitchatinov
   1993),  for a more
   reliable estimation of the $\alpha$ effect the
   hydrodynamic part is should also be incorporated.
   
   A debate is ongoing in the literature about whether the
   relation between $\alpha$ and    $h_c$,
    which was proposed by Keinig (1983), 
   R\"adler \& Seehafer (1990) and Seehafer (1994), is a
   fundamental one. It is a very important question because
   this relation can be used in solar physics as a diagnostic
   tool to for defining the $\alpha$ effect parameters in the
   Solar convection zone.  	
   The present computations show that the phase  relation
   between the two
    holds (in the sense Keinigs' formula) for the slow rotation case 
   eqs.(\ref{sa_sp},\ref{sh_sp},\ref{sa_d},\ref{sh_d}), 
   sf. R\"udiger \& Pipin (2000),
   and so does for the fast rotation case   eqs. (\ref{a_sp},\ref{h_sp},\ref{a_d},
   \ref{h_d}).
   The Keinig-Seehafer's formula is violated for intermediate 
   values of $\Omega^*$ and shear.
   This is clearly seen in Figures 3 and 4.
    The validity of Keinig-Seehafer's identity  depends largely
   on the assumed stationarity and homogeneity
   of the turbulent flows. Such an approximation is probably
    valid for the slow rotation case or for
   the saturated state in the fast rotation limit. In
   addition, 
   the phase relation can be violated 
   due to contributions from kinetic helicity. It is known that 
   the {\it isotropic} $\alpha$ effect can be expressed 
   by the sum of kinetic and current helicities, e.g.,
   Field et al (1999). Such a representation reflects the
   fact that $\alpha$ can be divided into the "hydrodynamic"
    and "magnetic" parts. We skip the "hydrodynamic" part in
    the paper. Hence we can not make a conclusive statement
    about the relation $\alpha$ to $h_c$ in a general case.
 
      Aplication of  $\alpha^{\phi\phi}$ contributed by shear
   to simple 1D model shows the
   drastic differences in evolution of the generated large-scale
   magnetic field compared to  the  case where
   $\alpha^{\phi\phi} \propto\cos\theta$.
   The shear contributions to $\alpha^{\phi\phi}$ cause 
   the spatial separation between
   $\alpha$ and $\Omega$ generation processes. As noted by
   Deinzer et al(1974), such a separation results  to increase
   of the dynamo period.  Our simulations show that the
   interface dynamo driven by shear-dominated $\alpha$ effect
   (and with parameters adjusted to match conditions near
   the bottom or right beneath the SCZ) has  a period about 1
    diffusion time which is comparable
   with the Solar cycle period. Though, the obtained butterfly
   diagrams are not in agreement with observations. This
   could be improve by including the contributions from the
   meridional circulation or from the anisotropic turbulent
   transport effects (Kitchatinov 1990). Unfortunately, at least for
   the solar case, the solution of
   the period cycle problem obtained from this 1D model can
   not be extended for the distributed dynamo in the
   spherical shell, because the radial shear is weak in the
   main part of convection zone and the Coriolis number is
   not strong enough there. This conclusion is well
   illustrated by Figures 2,3. As can be seen, the
   influence of the differential rotation on the latitudinal
   distribution of $\alpha$ is quite strong
   but not enough to be the decisive factor.
  
   Nevertheless, in the spherical shell dynamos the
   contributions  of the differential
   rotation to the $\alpha$ effect can introduce an
   interesting component in the non-linear feedback reaction of the
   magnetic field on the $\alpha$ effect. The magnetic field
   is well known to suppress the  $\alpha$. However, the
   numerical simulations show the radial gradient of the
   angular velocity is also likely
   to be increased in the regions with the strong magnetic
   fields (cf. Pipin 1999). Then we might well expect the
   $\alpha$ effect contributed by shear will be less
   suppressed by the Lorentz forces than contributions
   connected with rigid rotation.  This conclusion should
   be checked by computing the shear-dominated alpha effect
   after taking the influence of magnetic field on the
   turbulence into account. Such an argument also should be
   considered in numerical simulation devoted to
   the magnetic quenching  of alpha effect (cf. Ossendrijver et al 2001).

   Another interesting point which could be discussed in the
   light of the results obtained connects with idea that in fully developed
   turbulence the hydrodynamic and magnetic parts of
   $\alpha$ effect cancel each other, Vainshtein (1980). For the rigid
   rotation case, as we noted above, the
   magnetic part of alpha effect  is quenched opposite to
   increase the Coriolis number. However its hydrodynamic part tends
   to the constant (R\"udiger \& Kitchatinov 1993).
   Consequently,
   these counterparts are hardly canceled for the fast
   rotating system. Our paper (as well as paper of R\"udiger \&
   Kitchatinov 1993) is based on the considering
   the real sources of the alpha effect in rotating
   stratified turbulence. This differs from publications were
   the hydrodynamic and magnetic helicity spectrum are
   defined a priori
    (eg. Vainshtein 1980, Field et al. 1999). In
   principle, an ideal choice involves computing the combination
   of the hydrodynamic and magnetic parts of  the $\alpha$
   effect while including compressibility effects
   equally in  both the density and intensity stratifications
   cases and  further taking into account the differential
   rotation, and  the non-linear magnetic feedback. 
   We hope to take some steps along this direction in future.

   \section*{Acknowledgments}
I express my appreciation to RFBR grants No.
00-02-17854, No.00-15-96659, No 02-02-16044 and INTAS grant
No. INTAS-2001-0550. The large part of the work
presented in the paper was done during my visit to Indian
Institute of Astrophysics, Bangalore, funded by the Indian DST grant
No.INT/ILTP/FSHP-5/2001. I'm grateful to Prof. Cowsik and
Dr. Hiremath for hospitality. A massive tensor algebra
presented here was done with help of tensor manipulation package provided by
Maxima (a GPL descendant of DOE Macsyma, http://Maxima.sf.net).
 I would like to say many
thanks to participants of Maxima's mail list for help with
some technical questions about Maxima. 


\section{Appendix}
The functions used in this paper are as follows,
\begin{eqnarray*}
&&f_1=-\frac{1}{{4\Omega^*}^5}
\left({\Omega^*} \left( 3 + 2\,{\Omega^*}^2 \right)  - 3\left( 1 + {\Omega^*}^2 \right)\arctan
({\Omega^*})\right),\\
&&f_2=
\frac{3}{64\,{\Omega^*}^5}\bigl( {\Omega^*}\left( 15 + {\Omega^*}^2 \right)  +
      ( -15 - 6{\Omega^*}^2 +{\Omega^*}^4) \arctan ({\Omega^*}) \bigr),\\
&&f_4=\frac{1}{192{\Omega^*}^9( 1 + {\Omega^*}^2)}
\Bigl({\Omega^*}( -2205 - 3360{\Omega^*}^2 - 876{\Omega^*}^4 + 384{\Omega^*}^6 + 73{\Omega^*}^8) 
\\ 
&&+ 9( 245 + 455{\Omega^*}^2+ 200{\Omega^*}^4 
- 32{\Omega^*}^6 - 21{\Omega^*}^8 + {\Omega^*}^{10})\arctan ({\Omega^*})\Bigr),\\
&&f_6=\frac{1}{192{\Omega^*}^9( 1 + {\Omega^*}^2)} \bigl( {\Omega^*}(-2205 - 
3360{\Omega^*}^2  -1011{\Omega^*}^4+ 240{\Omega^*}^6 + 64{\Omega^*}^8 )\\
&& - 9(-245 -455{\Omega^*}^2- 215{\Omega^*}^4 + 11{\Omega^*}^6 + 
16{\Omega^*}^8)\arctan ({\Omega^*})\bigr),\\
&&f_7=\frac{1}{960{\Omega^*}^9( 1 + {\Omega^*}^2) }\Bigl(9135{\Omega^*} + 13440{\Omega^*}^3 
+ 3307{\Omega^*}^5 - 2148{\Omega^*}^7 - 870{\Omega^*}^9 \\
&&-15( 609 
+ 1099{\Omega^*}^2 + 465{\Omega^*}^4 - 121{\Omega^*}^6 - 
102{\Omega^*}^8 -6{\Omega^*}^{10})\arctan ({\Omega^*})\Bigr),\\
&&f_8=\frac{1}{960{\Omega^*}^9( 1 + {\Omega^*}^2)}
\Bigl({\Omega^*}( -9135 - 13440{\Omega^*}^2 
- 2002{\Omega^*}^4 + 4068{\Omega^*}^6 \\
&&+ 1485{\Omega^*}^8)+15(609 
+1099{\Omega^*}^2 + 378{\Omega^*}^4 - 278{\Omega^*}^6\\
&&-195{\Omega^*}^8- 29{\Omega^*}^{10})\arctan ({\Omega^*})\Bigr),\\
&&f_9=\frac{1}{960{\Omega^*}^9( 1 + {\Omega^*}^2)}
\Bigl(47565{\Omega^*} + 70560{\Omega^*}^3 + 15133{\Omega^*}^5 -
 12252{\Omega^*}^7 \\
&& - 3710{\Omega^*}^9 - 15(3171 + 5761{\Omega^*}^2 
  + 2295{\Omega^*}^4 - 751{\Omega^*}^6 - 486{\Omega^*}^8 \\
&&-  30{\Omega^*}^{10})\arctan ({\Omega^*})\Bigr),\\
&&f_{11}=\frac{1}{960{\Omega^*}^9( 1 + {\Omega^*}^2)}\Bigl(9135{\Omega^*} +
 13440{\Omega^*}^3 + 2182{\Omega^*}^5 - 3168{\Omega^*}^7 - 885{\Omega^*}^9 \\
&&+15( -609 - 1099{\Omega^*}^2 
 - 390{\Omega^*}^4 + 214{\Omega^*}^6 + 119{\Omega^*}^8 
 + 5{\Omega^*}^{10} )\arctan ({\Omega^*})\Bigr),\\
&&f_{12}=   \frac{1}{960{\Omega^*}^9( 1 + {\Omega^*}^2) }
\Bigl({\Omega^*}(-9135-13440{\Omega^*}^2 - 2587{\Omega^*}^4 + 2688{\Omega^*}^6 \\
&&+810{\Omega^*}^8 )  -15( -609 - 1099{\Omega^*}^2 
 - 417{\Omega^*}^4 + 173{\Omega^*}^6 + 110{\Omega^*}^8 \\
&& + 10{\Omega^*}^{10} )\arctan ({\Omega^*})\Bigr),\\
&&f_{13}= \frac{1}{64{\Omega^*}^5( 1 + {\Omega^*}^2)}\Bigl({\Omega^*}( -135 - 84{\Omega^*}^2 + 11{\Omega^*}^4)  + 
    ( 135 + 129{\Omega^*}^2 + 5{\Omega^*}^4 \\
 && + 11{\Omega^*}^6)\arctan ({\Omega^*})\Bigr),\\
&&f_{14}= \frac{1}{64{\Omega^*}^5(1 + {\Omega^*}^2) } 
\Bigl(-27{\Omega^*} + 4{\Omega^*}^3 + 7{\Omega^*}^5 + 
( 27 + 5{\Omega^*}^2 - 15{\Omega^*}^4 \\
&&+ 7{\Omega^*}^6)\arctan ({\Omega^*})\Bigr),\\
&&f_{15}=  \frac{1}{192{\Omega^*}^9(1 + {\Omega^*}^2) }\Bigl(2205{\Omega^*} + 
3360{\Omega^*}^3 + 1686{\Omega^*}^5 + 372{\Omega^*}^7 - 55{\Omega^*}^9 \\
&&+9( -245 - 455{\Omega^*}^2 
- 290{\Omega^*}^4 - 82{\Omega^*}^6 - {\Omega^*}^8 + {\Omega^*}^{10} ) \arctan ({\Omega^*})\Bigr),\\
&&f_{17}=\frac{1}{960{\Omega^*}^9( 1 + {\Omega^*}^2)^2}
\Bigl(63945{\Omega^*} + 143325{\Omega^*}^3 + 80954{\Omega^*}^5 - 17822{\Omega^*}^7\\ 
&&- 23331{\Omega^*}^9- 3615{\Omega^*}^{11}
+ 15( 1 + {\Omega^*}^2)^2( -4263 - 2450{\Omega^*}^2 \\
&&+ 960{\Omega^*}^4 + 570{\Omega^*}^6 + 15{\Omega^*}^8)\arctan ({\Omega^*})\Bigr)\\
&&f_{18}= \frac{1}{960{\Omega^*}^9( 1 + {\Omega^*}^2)}\Bigl(9135{\Omega^*} +
 13440{\Omega^*}^3 + 2182{\Omega^*}^5 - 2988{\Omega^*}^7 \\
&& - 825{\Omega^*}^9 +15( -609 - 1099{\Omega^*}^2 
- 390{\Omega^*}^4 + 202{\Omega^*}^6 + 111{\Omega^*}^8 \\
&&+ 9{\Omega^*}^{10})\arctan ({\Omega^*})\Bigr), \\
&&f_{19}= \frac{1}{192{\Omega^*}^9( 1 + {\Omega^*}^2)^2}
\Bigl(9135{\Omega^*} + 19635{\Omega^*}^3 + 10347{\Omega^*}^5 - 2369{\Omega^*}^7 \\
&&- 2634{\Omega^*}^9 - 354{\Omega^*}^{11} 
+15( 1 + {\Omega^*}^2)^2( -609 - 294{\Omega^*}^2 + 125{\Omega^*}^4 \\
&&+ 60{\Omega^*}^6 + 2{\Omega^*}^8)\arctan ({\Omega^*})\Bigr)\\
&&f_{20}= \frac{1}{32{\Omega^*}^5( 1 + {\Omega^*}^2)}
\Bigl(15{\Omega^*} + 4{\Omega^*}^3 - 3{\Omega^*}^5 - 
3( 5 + 3{\Omega^*}^2 - {\Omega^*}^4 \\
&&+ {\Omega^*}^6 )\arctan ({\Omega^*})\Bigr),  \\
&&f_{22}=\frac{1}{192{\Omega^*}^9( 1 + {\Omega^*}^2)}
\Bigl({\Omega^*}( 2205 + 3360{\Omega^*}^2 + 1686{\Omega^*}^4 + 804{\Omega^*}^6 \\
&&+ 281{\Omega^*}^8)-3( 735 + 1365{\Omega^*}^2 + 870{\Omega^*}^4 
+ 390{\Omega^*}^6 \\
&&+ 163{\Omega^*}^8 + 13{\Omega^*}^{10} )\arctan ({\Omega^*})\Bigr),\\
&&f_{23}=\frac{1}{192{\Omega^*}^9( 1 + {\Omega^*}^2 ) }\Bigl(2205{\Omega^*} + 3360{\Omega^*}^3 + 876{\Omega^*}^5 - 384{\Omega^*}^7 - 73{\Omega^*}^9 \\
&&-9(245 + 455{\Omega^*}^2 + 200{\Omega^*}^4 - 32{\Omega^*}^6 - 21{\Omega^*}^8 + {\Omega^*}^{10} )\arctan ({\Omega^*})\Bigr),\\
&&f_{25}=\frac{1}{4{\Omega^*}^5}\Bigl(-{\Omega^*}( 3 + 2{\Omega^*}^2)+3( 1 + {\Omega^*}^2)\arctan ({\Omega^*})\Bigr),\\
&&f_{27}=\frac{1}{4{\Omega^*}^5}\left(-3\,{\Omega^*} + \left( 3 + {\Omega^*}^2 \right) \,\arctan ({\Omega^*})\right),\\
&&f_{28}=\frac{1}{4{\Omega^*}^5}\left({\Omega^*}\,\left( 15 + 4\,{\Omega^*}^2 \right)  - 
3\,\left( 5 + 3\,{\Omega^*}^2 \right) \,\arctan ({\Omega^*})\right),\\
&&f_{29}=\frac{3}{64{\Omega^*}^5( 1 + {\Omega^*}^2)}\Bigl( {\Omega^*}( 75 + 68{\Omega^*}^2 + {\Omega^*}^4 ) 
      \\
&&  -(75+93{\Omega^*}^2 +17{\Omega^*}^4 -{\Omega^*}^6) \arctan ({\Omega^*}) \Bigr) \\
&&f_{30}= \frac{1}{96{\Omega^*}^9( 1 + {\Omega^*}^2)^2}
\Bigl(15435{\Omega^*} + 35175{\Omega^*}^3 + 22212{\Omega^*}^5 - 76{\Omega^*}^7 \\
&&- 2895{\Omega^*}^9- 283{\Omega^*}^{11} 
-3( 1 + {\Omega^*}^2)^2(5145 + 3150{\Omega^*}^2 - 590{\Omega^*}^4 \\
&&- 330{\Omega^*}^6 + 9{\Omega^*}^8)\arctan ({\Omega^*})\Bigr),\\
&&f_{31}= \frac{3}{64{\Omega^*}^5(1 + {\Omega^*}^2)}\Bigl({\Omega^*}(105 + 100{\Omega^*}^2 + 3{\Omega^*}^4)  + 
      3(-35 - 45{\Omega^*}^2 \\
&& - 9{\Omega^*}^4 + {\Omega^*}^6)\arctan ({\Omega^*}) \Bigr),\\
&&f_{32}= \frac{1}{32{\Omega^*}^5}\bigl(3{\Omega^*}( -1 
+ {\Omega^*}^2 )+( 3 - 2{\Omega^*}^2 + 3{\Omega^*}^4)\arctan ({\Omega^*}) \bigr),\\ 
&&f_{\alpha 1}=\frac{1}{160 {\Omega^*}^8( 1 + {\Omega^*}^2)}
\Bigl(10080{\Omega^*} + 15120{\Omega^*}^3 + 3431{\Omega^*}^5 - 2324{\Omega^*}^7 
- 595{\Omega^*}^9 \\
&&+15( -672 - 1232{\Omega^*}^2 - 505{\Omega^*}^4 + 137{\Omega^*}^6 +
 85{\Omega^*}^8 + 3{\Omega^*}^{10})\arctan ({\Omega^*})\Bigr),\\
&&f_{\alpha 2}=-\frac{1}{32\,{\Omega^*}^4}\left({\Omega^*}\,\left( 111 + 25\,{\Omega^*}^2 \right) 
 - \left( 111 + 62\,{\Omega^*}^2 + 7\,{\Omega^*}^4 \right) \,\arctan
 ({\Omega^*})\right),
\end{eqnarray*}
$f_{3,5,10,21}=-f_2,f_{16}=f_4,f_{24}=-\frac{3}{2}f_{20},
f_{26}=f_2,f_{33}= -f_{11}$. The 
current helicity was defined with the following functions, 
\begin{eqnarray*}
&&\psi_0=\frac{6\,{\Omega^*} + 4\,{\Omega^*}^3 - 6\,\left( 1 +
 {\Omega^*}^2 \right) \,\arctan ({\Omega^*})}{4{\Omega^*}^5},\\
&&\psi_1=\frac{-\left( {\Omega^*}\,\left( 33 + 58\,{\Omega^*}^2 +
 29\,{\Omega^*}^4 \right)  \right)  + 
    3\,{\left( 1 + {\Omega^*}^2 \right) }^2\,\left( 11 + 
    {\Omega^*}^2 \right) \,\arctan ({\Omega^*})}{48\,{\Omega^*}^5\,
    \left( 1 + {\Omega^*}^2 \right) },\\
&&\psi_2= \frac{}{96\,{\Omega^*}^5(1 + {\Omega^*}^2) }
\Bigl({\Omega^*}\,\left( -645 - 532\,{\Omega^*}^2 + 
9\,{\Omega^*}^4 \right) \\
&&+3\,\left( 215 + 249\,{\Omega^*}^2 + 37\,{\Omega^*}^4 + 
    3\,{\Omega^*}^6 \right) \,\arctan ({\Omega^*})\Bigr)  \\
&&\psi_3= \frac{1}{96{\Omega^*}^5( 1 + {\Omega^*}^2)}
\Bigl({\Omega^*}( 75 + 144{\Omega^*} + 92{\Omega^*}^2 + 240{\Omega^*}^3 
+ 9{\Omega^*}^4 + 96{\Omega^*}^5)  \\
&&+3( -25 - 48{\Omega^*} - 39{\Omega^*}^2 - 96{\Omega^*}^3 
- 11{\Omega^*}^4 - 48{\Omega^*}^5 + 3{\Omega^*}^6 )\arctan ({\Omega^*})\Bigr),\\
&&\psi_4= \frac{-6\,{\Omega^*} + 2\,\left( 3 + 
{\Omega^*}^2 \right) \,\arctan ({\Omega^*})}{4{\Omega^*}^5},\\    
&&\psi_5= \frac{2\,{\Omega^*}\,\left( 15 + {\Omega^*}^2 - 8\,{\Omega^*}^4 \right)  + 
    6\,\left( -5 - 2\,{\Omega^*}^2 + 3
    \,{\Omega^*}^4 \right) \,\arctan ({\Omega^*})}{12\,{\Omega^*}^5\,
    \left( 1 + {\Omega^*}^2 \right) },\\
&&\psi_6=\frac{{\Omega^*}\,\left( 27 + 54\,{\Omega^*}^2 + 23\,{\Omega^*}^4 \right)  - 
    9\,{\left( 1 + {\Omega^*}^2 \right) }^2\,\left( 3 + 
    {\Omega^*}^2 \right) \,\arctan ({\Omega^*})}{24\,{\Omega^*}^5\,
    \left( 1 + {\Omega^*}^2 \right) } ,\\
&&\psi_7\frac{{\Omega^*}\,\left( -15 - 4\,{\Omega^*}^2 + 3\,{\Omega^*}^4 \right)  + 
    3\,\left( 5 + 3\,{\Omega^*}^2 - {\Omega^*}^4 + 
    {\Omega^*}^6 \right) \,\arctan ({\Omega^*})}{48\,{\Omega^*}^5\,
    \left( 1 + {\Omega^*}^2 \right) } ,\\  
&&\psi_8= \frac{{\Omega^*}\,\left( 3 + 2\,{\Omega^*}^2 + 3\,{\Omega^*}^4 \right)  +
    3\,\left( -1 + {\Omega^*}^2 \right) \,{\left( 1 + 
    {\Omega^*}^2 \right) }^2\,\arctan ({\Omega^*})}{24\,{\Omega^*}^4\,
    \left( 1 + {\Omega^*}^2 \right) },\\
&&\psi_9=\frac{{\Omega^*}\,\left( 15 + 13\,{\Omega^*}^2 \right)  - 
    3\,\left( 5 + 6\,{\Omega^*}^2 + {\Omega^*}^4 \right)\arctan ({\Omega^*})}{24\,{\Omega^*}^5},\\
&&\psi_{10}= \frac{{\Omega^*}\,\left( -9 + {\Omega^*}^2 \right)  + 
\left( 9 + 2\,{\Omega^*}^2 + {\Omega^*}^4 \right) \,\arctan ({\Omega^*})}
  {16\,{\Omega^*}^4},\\                 
&&\psi_{h1}=\frac{1}{96{\Omega^*}^4( 1 + {\Omega^*}^2)}\Bigl({\Omega^*}
( -405 - 228{\Omega^*}^2 + 73{\Omega^*}^4) \\
&&+3(135 + 121{\Omega^*}^2 - 11{\Omega^*}^4 + 3{\Omega^*}^6)
\arctan ({\Omega^*})\Bigr),\\
&&\psi_{h2}=\frac{1}{96\,{\Omega^*}^4\,
    \left( 1 + {\Omega^*}^2 \right) }\Bigl({\Omega^*}\,\left( 117 + 
192\,{\Omega^*}^2 + 43\,{\Omega^*}^4 \right)  \\
&&- 3\,\left( 39 + 77\,{\Omega^*}^2 + 45\,{\Omega^*}^4 +
     7\,{\Omega^*}^6 \right) \,\arctan ({\Omega^*})\Bigr)
\end{eqnarray*}

\end{document}